\documentclass[prl,onecolumn,amsmath,amssymb,superscriptaddress]{revtex4-1}
\usepackage{bm,psfrag,graphicx,setspace}
\usepackage{braket}
\usepackage[left]{lineno}
\usepackage{xcolor}
\selectfont

\DeclareFontFamily{OMS}{oasy}{\skewchar\font48 }
\DeclareFontShape{OMS}{oasy}{m}{n}{%
         <-5.5> oasy5     <5.5-6.5> oasy6
      <6.5-7.5> oasy7     <7.5-8.5> oasy8
      <8.5-9.5> oasy9     <9.5->  oasy10
      }{}
\DeclareFontShape{OMS}{oasy}{b}{n}{%
       <-6> oabsy5
      <6-8> oabsy7
      <8->  oabsy10
      }{}
\DeclareSymbolFont{oasy}{OMS}{oasy}{m}{n}
\SetSymbolFont{oasy}{bold}{OMS}{oasy}{b}{n}

\DeclareMathSymbol{\smallleftarrow}     {\mathrel}{oasy}{"20}
\DeclareMathSymbol{\smallrightarrow}    {\mathrel}{oasy}{"21}
\DeclareMathSymbol{\smallleftrightarrow}{\mathrel}{oasy}{"24}

\begin{document}

\author{Marc R. Bourgeois}
\affiliation
{Department of Chemistry, University of Washington, Seattle WA, 98195}
\author{Andrew W. Rossi}
\affiliation
{Department of Chemistry, University of Washington, Seattle WA, 98195}
\author{David J. Masiello}
\email{masiello@uw.edu}
\affiliation
{Department of Chemistry, University of Washington, Seattle WA, 98195}

\title{Strategy for Direct Detection of Chiral Phonons with Phase-Structured Free Electrons}




\begin{abstract}
Chiral phonons possessing valley pseudo angular momentum (PAM) underlie a diversity of quantum phenomena of fundamental and applied importance, but are challenging to probe directly. We show that deficiencies of typical momentum-resolved electron energy loss measurements that make it impossible to distinguish the PAM of chiral phonons can be overcome by introducing pinwheel free electron states with well-defined PAM. Transitions between such states generate 2D periodic arrays of in-plane field vortices with polarization textures tailored to selectively couple to desired chiral mode symmetries.
\end{abstract}


\maketitle


Chirality plays a central role in nearly all aspects of physics, chemistry, and biology. At large length scales, chiral gravitational waves have been considered as sources of circularly polarized contributions to the cosmic microwave background \cite{inomata2019chiral}, while at short length scales chirality dictates molecular interactions and organizations underpinning biological processes. In condensed matter physics, chirality and spin-momentum locking are at the heart of the menagerie of identified Hall effects, as well as stable \cite{al2001skyrmions,yu2018transformation} and transient free space \cite{gao2020paraxial, shen2021supertoroidal} and evanescent \cite{tsesses2018optical, du2019deep, dai2020plasmonic, davis2020ultrafast, ghosh2023spin} topological electromagnetic field textures. Following the report \cite{chen2015helicity} of helicity-resolved Raman scattering experiments from transition-metal dichalcogenide (TMD) atomic layers, it was predicted \cite{zhang2015chiral} that two-dimensional (2D) atomic crystals with hexagonal symmetry can host chiral phonons with well-defined pseudo angular momentum (PAM) arising from discrete rotational crystal symmetry. Direct experimental evidence of chiral phonons was subsequently reported based on transient infrared circular dichroism measurements of a monolayer TMD \cite{zhu2018observation}, generating significant interest in the potential for leveraging chiral phonons to achieve novel material properties that are sensitive to low-energy spin-selective excitations. Beyond 2D hexagonal lattices, chiral phonons have been identified in other 2D \cite{chen2018chiral, chen2019chiral, suri2021chiral, wang2022chiral, zhang2022chiral, fransson2023chiral} and 3D \cite{chen2021propagating, ishito2023truly} crystalline systems, as well as in bio-organic molecules \cite{ueda2023chiral}. Chiral phonons have also been recently identified as participants in various exotic condensed matter processes including anomalous thermal Hall conductivity \cite{grissonnanche2020chiral} and the chiral-phonon-activated spin Seebeck effect \cite{kim2023chiral}, among others \cite{jeong2022unconventional, chen2022chiral}. Despite the high-level of current interest in chiral phonons, their experimental detection has been largely limited to optical Raman scattering measurements, with spatial resolution and linear momentum limited by the photon wavelength, and direct detection remains an outstanding challenge impeding further progress \cite{wang2024chiral}.

Meanwhile, instrumental advances during the past decade in aberration correction, monochromation, and detector technologies have made it possible to perform vibrational spectroscopy inside a scanning transmission electron microscope (STEM) \cite{krivanek2014vibrational, dwyer2016electron, lagos2017mapping, hage2018nanoscale, hachtel2019identification}. In particular, momentum-resolved electron energy loss ($q$-EEL) spectroscopy \cite{dwyer2016electron, senga2019position, hong2020probing, ohara2023high}, and the related high-resolution EEL spectroscopy (HREELS) technique \cite{li2023direct, li2024observation}, have been used to measure the energy-momentum dispersion of phonons in mono- and few-layer 2D atomic crystals. While these measurements have demonstrated the ability of inelastic free electron scattering to complement other phonon measurement techniques, such as inelastic neutron and x-ray scattering, the absence of an obvious electron polarization degree of freedom has hindered the spin-selective detection of chiral phonons using free electrons. However, the transverse phase shaping and coherent detection schemes that were originally developed for core-loss EEL spectroscopy and that enabled magnetic circular dichroism measurements in the TEM \cite{Yuan1997-hj, Hebert2003-mn, Schattschneider2006-fy} have been recently adapted to the low loss regime. By selectively monitoring inelastic scattering of free electrons between states with tailored transverse phase profiles, energy loss processes involving desired target mode symmetries \cite{guzzinati2017probing} and OAM states \cite{asenjo2014dichroism, zanfrognini2019orbital} have been demonstrated, and optical polarization analogs in free electron scattering have been identified \cite{lourencco2021optical, bourgeois2023optical}. The full potential of these measurement schemes is currently being explored, while the transverse phase shaping and selection technologies underlying them are actively under development \textcolor{black}{\cite{niermann2014creating}}\cite{yu2023quantum, PhysRevResearch.2.043227, Madan2022-gj, harvey2014efficient, Tsesses2023-hw}. 

In this Letter we provide a strategy for leveraging inelastic scattering of transversely phase shaped free electrons to achieve direct detection of chiral phonons within a STEM. We introduce pinwheel free electron states composed of a small number of phased plane wave components that share the same reciprocal space three-fold discrete rotational symmetry as chiral phonons in hexagonal monolayer 2D atomic crystals at the $\textrm{K}$ and $\textrm{K}'$ points, which introduces a new PAM degree of freedom to the probing electrons. In analogy to previous work on OAM-resolved EEL spectroscopy \cite{lloyd2012quantized, asenjo2014dichroism, cai2018efficient, zanfrognini2019orbital, lourencco2021optical, bourgeois2022polarization, bourgeois2023optical}, pre- and post-selection of the transverse free electron states enables tracking of PAM exchange between pinwheel free electrons and the sample. We trace the utility of the identified phase-shaped free electron transitions to the topological spin textures of their associated transition electric fields, which can be tailored to selectively couple to chiral phonons of the same symmetry \textcolor{black}{and PAM}. Compared to light-based probes, inelastic electron scattering offers \textcolor{black}{the potential for} superior spatial resolution, in addition to an often simpler interpretation since the Brillouin zone (BZ) edges can be accessed directly. This work informs current and future efforts to detect and characterize chiral phonons using inelastic free electron scattering, and identifies specific experimental geometries and parameters to achieve detection.

\begin{figure}
    \centering
    \includegraphics{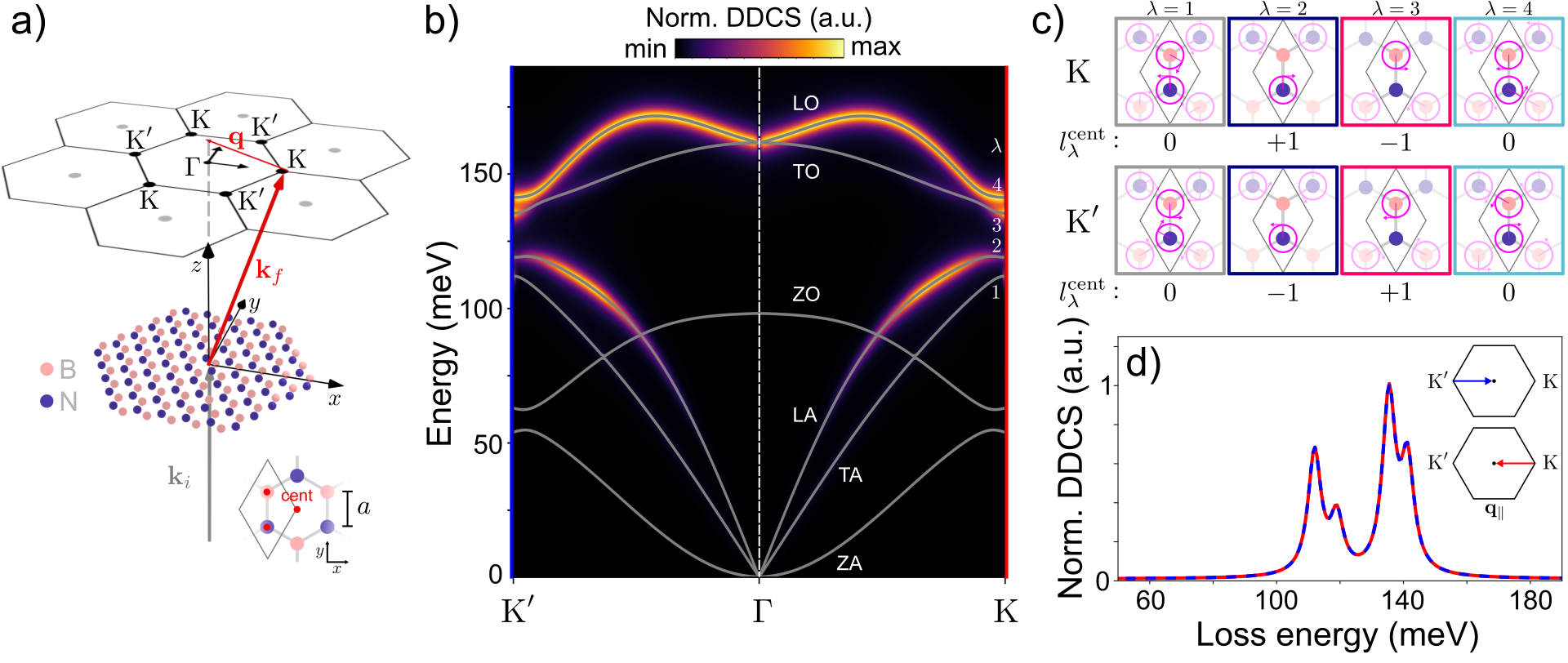}
    \caption{hBN phonons and $q$-EEL spectroscopy. a) Scheme of $q$-EEL process involving momentum transfer between initial $\mathbf{k}_i$ (gray) and final $\mathbf{k}_f$ (red) free electron states. The recoil momentum $\hbar{\bf q}$ has projection ${\bf q}_\parallel=({\bf q}\cdot\hat{\bf x})\hat{\bf x}$ onto the 2D sample situated in the $xy$ plane. \textcolor{black}{The three $C_3$ axes ($C_3^{\text{B}}$, $C_3^{\text{N}}$, and $C_3^{\text{cent}}$) associated with hBN are indicated by red markers in the lower inset.} b) Calculated hBN phonon eigenmode dispersion (gray) along $\textrm{K}' - \Gamma - \textrm{K}$ and normalized DDCS (color map). c) Real-space in-plane phonon eigenmodes at $\textrm{K}$ and $\textrm{K}'$. Pink circles indicate circular motion of sublattices, while arrows show the direction of rotation. The PAM \textcolor{black}{$l^{\text{cent}}_{\lambda}$} is noted below each panel \textcolor{black}{with respect to $C_3^{\text{cent}}$}. d) Lineouts of the DDCS spectra at $\textrm{K}$ and $\textrm{K}'$ from panel (b). All calculations include an empirical damping rate of $\eta=5$ $\textrm{meV}/\hbar$ at zero temperature and a vacuum background. }
    \label{F1}
\end{figure}


As a model 2D atomic crystal with hexagonal symmetry \textcolor{black}{that lacks inversion symmetry}, we consider chiral phonons in monolayer hexagonal boron nitride (hBN) situated in the $xy$ plane as depicted in Fig. \ref{F1}a. Eigenenergies $\hbar \omega_{\lambda}(\mathbf{k}_{\parallel})$ and eigenvectors $\boldsymbol{\xi}_\lambda(\mathbf{k}_{\parallel})$ of the six hBN phonon bands, indexed by $\lambda$, are computed within the harmonic crystal approximation \cite{dove1993lattice} with force constants taken from the literature \cite{bosak2007elasticity, michel2008theory, michel2009theory}. The gray lines in Fig. \ref{F1}b show the eigenenergies as a function of $\mathbf{k}_\parallel$ within the first BZ along $\textrm{K}' - \Gamma - \textrm{K}$. In contrast to graphene, the lack of inversion symmetry within the 2D plane and the fact that time reversal symmetry is broken at the $\textrm{K}$ and $\textrm{K}'$ points leads to non-degenerate in-plane valley phonon modes at the $\textrm{K}$ and $\textrm{K}'$ reciprocal space points with well-defined PAM \cite{zhang2015chiral}. Real-space motion of the four in-plane phonon eigenmodes \textcolor{black}{($\lambda = 1-4$)}, at the $\textrm{K}$ and $\textrm{K}'$ points are shown in Fig. \ref{F1}c. \textcolor{black}{Below each image is listed the total PAM $l_{\lambda}^{\text{cent}}$ with respect to the $C_3^{\text{cent}}$ rotation axis marked in red at the center of the hexagon shown in the panel (a) inset. Additional details regarding the PAM of in-plane phonons, including definition of PAM relative to other $C_3$ axes, is detailed in \cite{SM_chiralphonon}.} The lowest $(\lambda = 1)$ and highest $(\lambda = 4)$ energy in-plane modes involve opposing circular motion on both sublattices, with total PAM \textcolor{black}{$l_{\lambda}^{\text{cent}} = 0$}. For the other two in-plane modes \textcolor{black}{$(\lambda = 2,3)$ }, while one sublattice remains motionless, the other executes circular motion, the handedness of which ultimately determines whether the PAM \textcolor{black}{$l_{\lambda}^{\text{cent}} = \pm 1$}. The depicted phonons exhibit spin-momentum locking in that the $z$-oriented spin associated with each sublattice (equivalent to handedness of circular motion) changes sign under linear momentum reversal, i.e., $\textrm{K}\leftrightarrow \textrm{K}' $.


For an incoming probing electron with speed $v_i$, the state-resolved loss rate describing inelastic scattering at first order from the 2D atomic crystal at loss energy $\hbar \omega_{if}$ is \cite{hohenester2018inelastic}
\begin{equation}
    \begin{aligned}
        w_{fi}^{\textrm{loss}} &= \frac{2}{\hbar} \Big(\frac{v_i}{L}\Big)^2 \frac{L^3}{\Omega_0}\,\textrm{Im} \sum_{\lambda} \sum^{\textrm{BZ}}_{\mathbf{k}} \frac{  |\tilde{\mathcal{F}}_{\lambda}(\mathbf{k} )|^2 }{\omega_{\lambda}^2-\omega_{if}(\omega_{if}+ i\eta)}, \\
    \end{aligned}
    \label{wfi_planewave_sup}
\end{equation}
where $L^3$ defines the probing electron's quantization volume, $\Omega_0$ is the unit cell volume, and $\mathbf{k}$ is within the first BZ. The quantity $\tilde{\mathcal{F}}_{\lambda}(\mathbf{k} ) = \sum_{\mathbf{G}} \sum_{\kappa}\boldsymbol{\xi}^*_{\lambda\kappa}\cdot M_{\kappa}^{-1/2}\cdot e\bar{\mathbf{Z}}_{\kappa}(\mathbf{k}+\mathbf{G})\cdot \mathbf{E}_{fi,\mathbf{k}+ \mathbf{G} \kappa}^{0}$ involves the atomic masses $M_\kappa$ and effective charges $eZ_\kappa$ of each sublattice, indexed by $\kappa$, and $\mathbf{G}$ denotes a reciprocal lattice vector. $\mathbf{E}_{fi,\mathbf{k}+ \mathbf{G} \kappa}^{0}$ is the Fourier coefficient of the quasistatic transition electric field $\mathbf{E}^0_{fi}(\mathbf{k}) =-i \mathbf{k} (4 \pi /k^2) \, \rho_{fi}(\mathbf{k})$, where $\rho_{fi}(\mathbf{k})$ is the Fourier transform of the charge density $\rho_{fi}(\mathbf{x}) = -e \psi^*_{f}(\mathbf{x}) \psi_{i}(\mathbf{x})$ associated with the free electron transition from initial state $\psi_{i}(\mathbf{x})$ to final state $\psi_{f}(\mathbf{x})$. A phenomenological damping parameter $\eta$ is introduced to give spectral features finite width. 

Fig. \ref{F1}a shows a generic wide field inelastic electron scattering event as typically measured by $q$-EELS whereby an incoming free electron plane wave with wave vector $\mathbf{k}_i$ (gray) scatters to an outgoing plane wave with wave vector $\mathbf{k}_f$ (red) via interaction with the atomic crystal. In the low loss energy range considered, the linear momentum $\hbar \mathbf{q} = \hbar (\mathbf{k}_i - \mathbf{k}_f)$ lost by the probing electron is transferred into excited phonons with Bloch vector $\mathbf{k}_{\parallel}$.
The transition field associated with this situation is \cite{SM_chiralphonon}
\begin{equation}
    \mathbf{E}^0_{\mathbf{q},\mathbf{k}+\mathbf{G} \kappa} = \frac{4 \pi ie}{L^3}\frac{L}{v_i} \frac{\mathbf{k} + \mathbf{G}}{|\mathbf{k} + \mathbf{G}|^2} \, e^{i (\mathbf{k} + \mathbf{G})\cdot \mathbf{r}_\kappa}\delta_{\mathbf{k} + \mathbf{G}, \mathbf{q}},
    \label{Efi_pw}
\end{equation}
which is linearly polarized along the recoil direction $\hat{\mathbf{q}}$. The rate of phonon mode excitation is dictated by $\tilde{\mathcal{F}}_{\lambda}(\mathbf{q} ) \propto \boldsymbol{\xi}^*_{\lambda\kappa} \cdot \mathbf{E}^0_{\mathbf{q},\mathbf{k}+\mathbf{G} \kappa} \propto \boldsymbol{\xi}^*_{\lambda\kappa} \cdot \hat{\mathbf{q}} $. Thus, the linearly polarized transition field couples identically to circularly polarized in-plane phonons, and it is not possible to distinguish the PAM of the phonon modes at $\textrm{K}$ and $\textrm{K}'$ points from $q$-EEL spectra alone. Indeed, spectral lineouts of the double differential cross section (DDCS) \cite{SakuraiModern, rossi2024probing} at the $\textrm{K}$ and $\textrm{K}'$ points from Fig. \ref{F1}b  are shown in Fig. \ref{F1}d to be indistinguishable.


\begin{figure}
    \centering
    \includegraphics{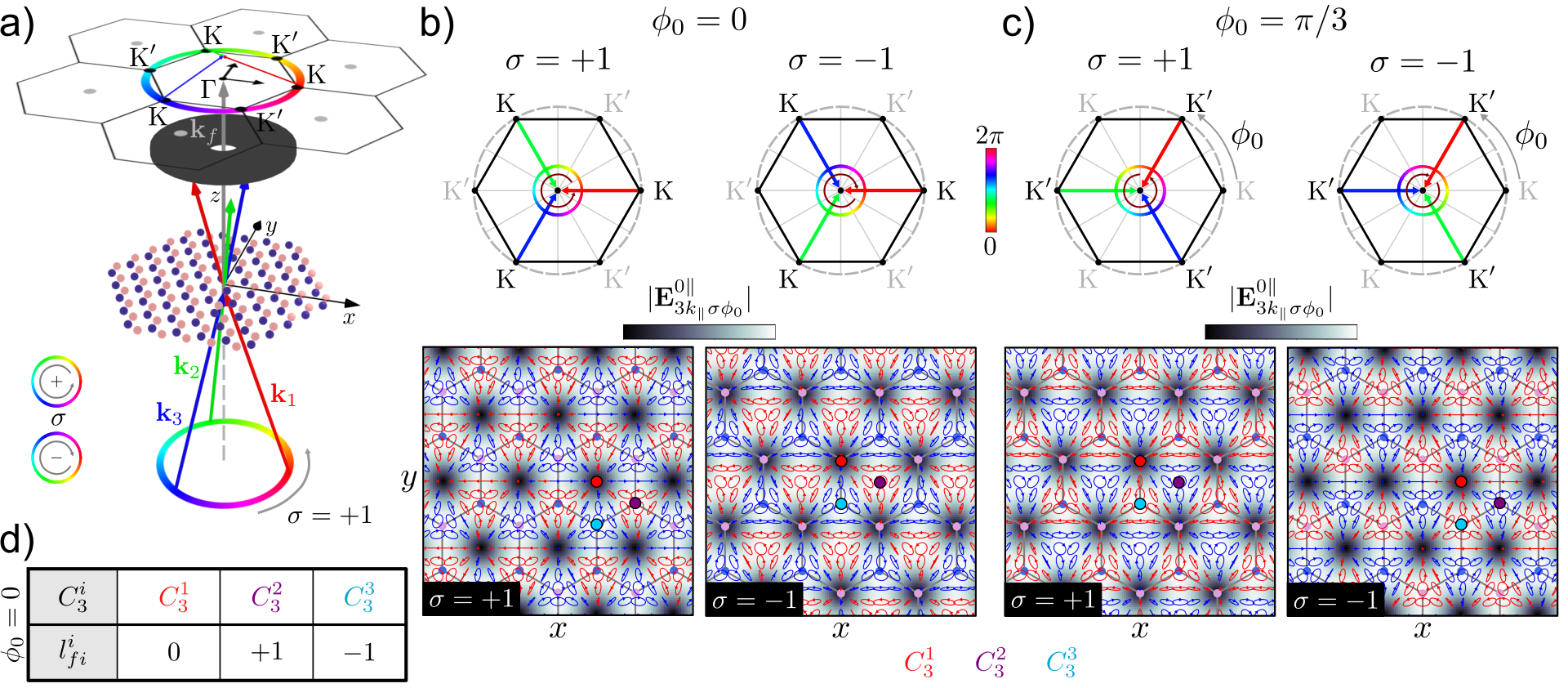}
    \caption{Transition electric fields involving three-fold pinwheel states. a) Scheme showing a transition from the $|P=3 , k_\parallel , \sigma = +1 , \phi_0 \rangle$ pinwheel state to $|\mathbf{k}^{\parallel}_f = \mathbf{0}\rangle$. b) and c) show in-plane magnitudes (color maps) and polarization textures (polarization ellipses) of $\mathbf{E}_{3 k_\parallel \sigma \phi_0 }^{0 \parallel}(\mathbf{x}_\parallel,z=0)$ associated with transitions $|P=3 , k_\parallel = \textrm{K} , \sigma = \pm1 , \phi_0 \rangle \to |\mathbf{k}^{\parallel}_f = \mathbf{0}\rangle$ at $\phi_0 = 0$ and $\phi_0 = \pi/3$, respectively. Red (blue) polarization ellipses indicate anticlockwise (clockwise) elliptical polarization rotation, and $\mathbf{R}_0 = -a/2 \  \hat{\mathbf{y}}$. \textcolor{black}{d) Total PAM $l_{fi}^{i}$ of the transition field relative to the $C_3^{i}$ $(i = 1,2,3)$ rotation centers marked in panel (b) for $\phi_0 = 0$.}  }
    \label{F2}
\end{figure}

It is perhaps natural to consider employing prototypical free electron states that carry well-defined units of intrinsic OAM, e.g., Bessel and Laguerre-Gauss beams \cite{uchida2010generation, verbeeck2010production, bliokh2017theory}. Previous works have demonstrated the ability to monitor OAM exchange between target sample and probing electron through careful pre- and post-selection of the OAM state of the free electrons \cite{lloyd2012quantized, asenjo2014dichroism, cai2018efficient, zanfrognini2019orbital, lourencco2021optical, bourgeois2022polarization, bourgeois2023optical}. However, this strategy is also problematic in the present case, since the reciprocal space wave function densities of these free electron states are azimuthally symmetric, leading to the inability to selectively probe the $\textrm{K}$ or $\textrm{K}'$ points. To circumvent this challenge, we introduce pinwheel initial states $|P\, k_\parallel\, \sigma\, \phi_0 \rangle$ of the free electron probe defined as 
\begin{equation}
    \psi_{P k_\parallel \sigma\phi_0}(\mathbf{x}) = \frac{1}{\sqrt{P}} \sum_{j=1}^P e^{\sigma i \phi_j} L^{-3/2} e^{i \mathbf{k}_j \cdot \mathbf{x}},
\end{equation}
which are composed of $P$ phased plane wave components with identical in-plane projection $k_\parallel = |\mathbf{k}_\parallel|$ and phases $\phi_j = (2 \pi/P) j + \phi_0$ winding either counterclockwise ($\sigma=+1$) or clockwise ($\sigma=-1$). The constituent wave vectors for such an incident state are $\mathbf{k}^i_j = ( k^i_\parallel \cos \phi_j, \, k^i_\parallel \sin \phi_j, \, k^i_z)$, with $k_i^2 = k^{i2}_z + k^{i2}_\parallel$. For 60 keV probing electrons typical of STEM EEL spectroscopy, the plane waves contributing to the pinwheel superposition states require convergence angles of $\theta_c \sim 13.5$ mrad for $k_\parallel$ at the hBN $\textrm{K}/\textrm{K}'$ points, while relative phases may be imprinted using techniques to shape transverse free electron wave functions currently under development \textcolor{black}{\cite{niermann2014creating}}\cite{yu2023quantum, PhysRevResearch.2.043227, Madan2022-gj, harvey2014efficient, Tsesses2023-hw}. 


By preparing an incoming pinwheel state and post-selecting the on-axis plane wave state with $\mathbf{k}^f_{\parallel} = {\bf 0}$, as depicted in Fig. \ref{F2}a using a spatial filter in the Fourier plane, the measured inelastic scattering process will be composed of $P$ phased momentum recoils with $\mathbf{q}_{j \parallel} = \mathbf{k}^i_{j \parallel}$, and the transition field becomes \cite{SM_chiralphonon}
\begin{equation}
    \begin{split}
        \mathbf{E}^0_{P k_\parallel \sigma\phi_0, \mathbf{k} + \mathbf{G} \kappa} = \frac{4 \pi ie }{L^3} \frac{L}{v_i} \frac{1}{\sqrt{P}} \sum_{j=1}^P \frac{ \mathbf{q}_j}{|\mathbf{q}_j|^2}e^{\sigma i \phi_j} e^{i (\mathbf{k} + \mathbf{G}) \cdot (\mathbf{r}_\kappa - \mathbf{R}_0)} \delta_{ \mathbf{k}+ \mathbf{G}, \mathbf{q}_j}.  \\
    \end{split}
    \label{Efi_pinwheel}
\end{equation}
Here, $\mathbf{R}_0$ dictates the displacement of the transition field interference pattern in the $xy$ plane relative to some arbitrarily placed origin\textcolor{black}{, which we take as the center of a real-space unit cell}. In analogy to topological electromagnetic vorticies created by interfering free space \cite{gao2020paraxial, shen2021supertoroidal} and evanescent \cite{tsesses2018optical, du2019deep, dai2020plasmonic, davis2020ultrafast, ghosh2023spin} electromagnetic waves with controlled geometric phase, the in-plane polarization of the transition field $\mathbf{E}^{0\parallel}_{Pk_\parallel\sigma\phi_0}(\mathbf{x})$ consists of a 2D periodic array of field vorticies. The presently discussed topological field textures are distinguished from those previously reported in that the transition field is sourced by the charge density associated with the probing free electron transition, leading to field texture periodicity on the $\sim$ {\AA} length scale. For the case $P=3$ and $k_\parallel = \textrm{K,\,K}'$, the phased momentum transfers have the same three-fold rotational reciprocal space symmetry as the $\textrm{K}$ and $\textrm{K}'$ phonons \cite{zhang2015chiral, zhu2018observation}, and the real space polarization array has hexagonal symmetry with unit cell dimensions matching those of the atomic crystal as presented in Fig. \ref{F2}b for $\phi_0 = 0$. Superimposed above the 2D $|\mathbf{E}^{0\parallel}_{3 k_\parallel \sigma \phi_0}(\mathbf{x}_\parallel, z=0)|$ maps are polarization ellipses colored red (blue) to indicate anticlockwise (clockwise) polarization rotation. \textcolor{black}{In close analogy to the situation encountered for valley phonons in 2D atomic crystals with hexagonal symmetry, the in-plane transition field has well-defined PAM $l_{fi}^{i} \in \{-1, 0, +1\}$ relative to any of the three distinct $C_3^{i}$ $(i = 1,2,3)$ rotation centers marked with colored dots in Figs. \ref{F2}b,c \cite{SM_chiralphonon}.} Reversing the winding direction of the plane wave phases by $\sigma = +1 \leftrightarrow \sigma=-1$ leads to a spatial shift of the field nodes relative to the atomic crystal sites. \textcolor{black}{Consequently, $l_{fi}^{i}$ does not depend on the value of $\sigma$, and $l_{fi}^{i}$ is tabulated in Fig. \ref{F2}d for $\phi_0 = 0$.} The transition fields shown in Fig. 2b,c exhibit spin-momentum locking in the sense that simultaneous exchange of $\sigma=+1 \leftrightarrow \sigma=-1$ and $\phi_0 = 0 \leftrightarrow \phi_0 = \pi/3$, which is equivalent to $\textrm{K} \leftrightarrow \textrm{K}'$, yields the same spatial in-plane field intensity profile, with regions of clockwise and anticlockwise polarization rotation interchanged. \textcolor{black}{This property reflects the fact that $l^{i}_{fi}  \to - l^{i}_{fi}$ when $\phi_0=0 \to \pi/3$ \cite{SM_chiralphonon}.}


\begin{figure}
    \centering
    \includegraphics{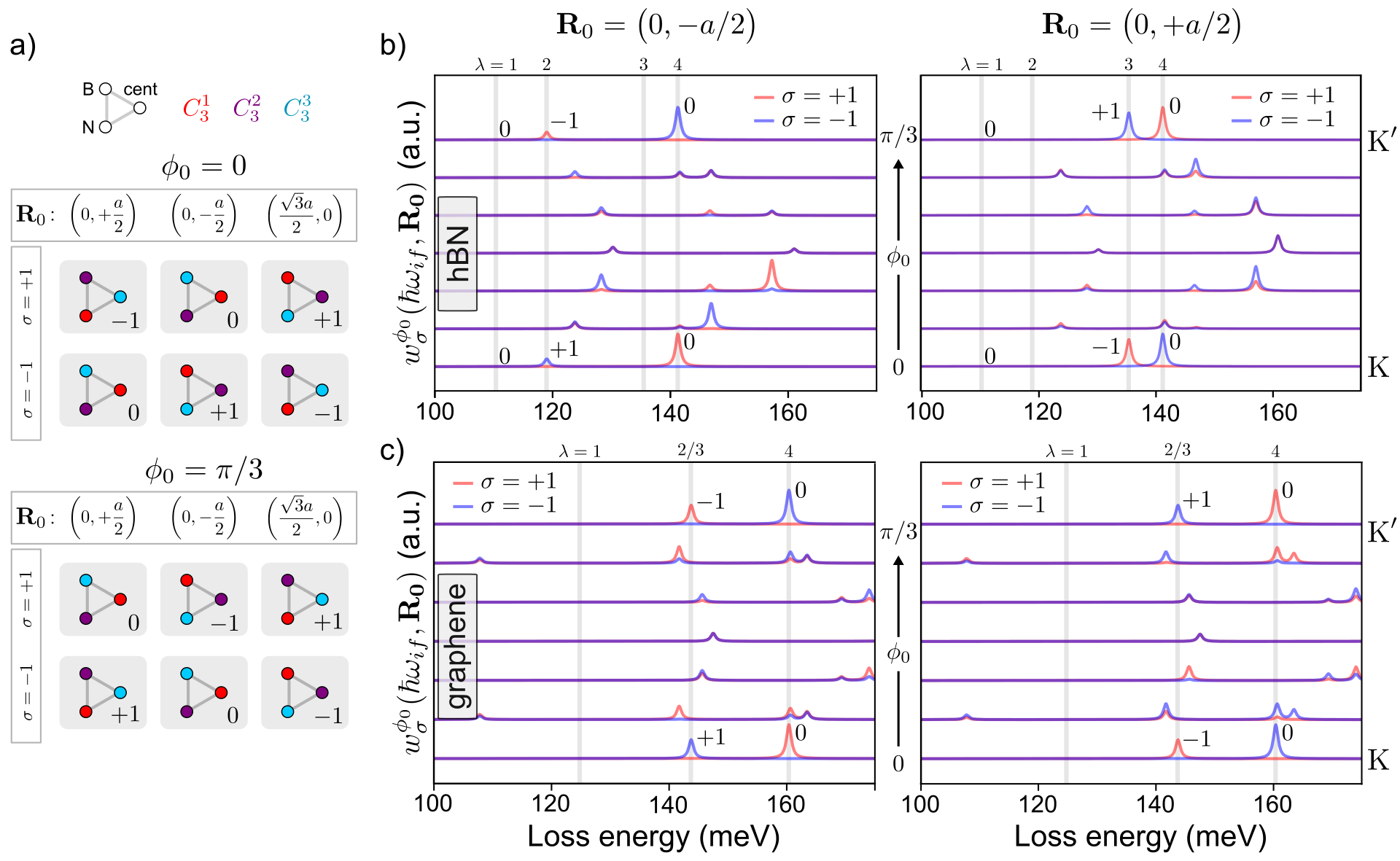}
    \caption{\textcolor{black}{Chiral phonon discrimination with phase shaped electrons. a) Configuration tables specifying the identities of co-located atomic crystal and transition field $C_3$ axes for $\mathbf{R}_0 =  \big\{ \big(0, \pm a/2 \big), \big(\sqrt{3} a/2, 0 \big) \big\}$, $\sigma = \pm 1$, and $\phi_0 = \{ 0, \pi/3 \}$. Circles at the vertices of each triangle represent B, N, and center (cent) atomic crystal positions, while circle color indicates which of the transition field $C_3^{i}$ axes resides at each site. The integer to the lower right of each diagram indicates $l^{\text{cent}}_{fi}$. b) Calculated transition rates $w^{\phi_0}_{\sigma}(\hbar \omega_{if}, \mathbf{R}_0)$ for incoming pinwheel states transitioning to the on-axis plane wave state as a function of loss energy and $ 0 \leq \phi_0 \leq \pi/3$ for an hBN monolayer with $\mathbf{R}_0=\big( 0, \pm a/2\big)$. Red (blue) traces indicate $\sigma=+1$ ($\sigma=-1$). Integers adjacent to spectral features at $\text{K}/\text{K}'$ indicate the $l^{\text{cent}}_{fi}$ from (a) relevant to each peak. c) Same as (b), but for monolayer graphene. All calculations assume a vacuum background environment and zero temperature. An empirical phonon damping rate of $\eta=1$ $\textrm{meV}/\hbar$ is included.}}
    \label{F3}
\end{figure}


\textcolor{black}{
The ability of the identified free electron transitions to generate electric fields that mirror the spin texture and spin-momentum locking of the hBN chiral phonons makes it possible to leverage the former to directly probe the latter. In particular, spatially scanning the transition field relative to the atomic crystal by changing $\mathbf{R}_0$ leads to either alignment, or varying degrees of misalignment, of the three $C_3$ axes of the transition field with those of the atomic crystal. When the two groups of $C_3$ axes are aligned, the PAM of both the transition field and the phonons can be simultaneously well-defined relative to a common origin, and coupling occurs between transition and phonon displacement fields with equal PAM as required by PAM conservation. Conversely, when the transition field and atomic crystal $C_3$ axes are misaligned, the PAMs of the phonon branches cease to be well-defined relative to the $C_3$ axis of the probe, and exchange of intrinsic and extrinsic PAM occurs. 
}

\textcolor{black}{
When $\mathbf{R}_0 \in \{ (0, \pm a/2), (\sqrt{3}a/2,0) \}$, the transition field $C_3^{i}$ and atomic crystal rotation center positions coincide, with the particular value of $i$ at each site dictated by the other parameters characterizing the transition field, i.e., $\sigma=\pm 1$ and $\phi_0 \in \{ 0, \pi/3\}$. The possible configurations of co-located $C_3$ axes are pictorially represented in Fig. \ref{F3}a. For example, when $\mathbf{R}_0 = (0,-a/2)$, $\sigma = +1$, and $\phi_0 = 0$, it is the $C_3^{1}$ transition field axis at the location of the $C_3^{\text{cent}}$ atomic crystal rotation center. Therefore, since $l^{1}_{fi} = 0$, this transition field couples only to valley phonons with PAM $l^{\text{cent}}_\lambda = 0$. However, for $\mathbf{R}_0 = (0,-a/2)$, $\sigma = -1$, and $\phi_0 = 0$, $C_3^{2}$ coincides with $C_3^{\text{cent}}$, and phonons with PAM $l^{\text{cent}}_\lambda = +1$ are excited. This is precisely the behavior exhibited by the corresponding state- and energy-resolved transition rates $w^{\phi_0}_{\sigma}(\hbar \omega_{if}, \mathbf{R}_0)$ presented in Fig. \ref{F3}b for hBN, which are evaluated using Eqs. \eqref{wfi_planewave_sup} and \eqref{Efi_pinwheel}. Red and blue traces correspond to $\sigma = +1$ and $\sigma = -1$, respectively. For $\mathbf{R}_0 = (0,-a/2)$ and $\phi_0 = 0$, the red trace shows features at the energies of modes $\lambda = 1$ and $\lambda = 4$, which have PAM $l^{\text{cent}}_\lambda = 0$, while the blue trace exhibits a single peak at mode $\lambda = 2$ with $l^{\text{cent}}_\lambda = +1$. Meanwhile, for $\mathbf{R}_0 = (0,+a/2)$ and $\phi_0 = 0$, in agreement with PAM conservation considerations encoded into the table in Fig. \ref{F3}a, the $\sigma = +1$ and $\sigma = -1$ transition fields couple to valley phonons with PAM $l^{\text{cent}}_\lambda = -1$ and $l^{\text{cent}}_\lambda = 0$, respectively. Using $l^{\text{cent}}_{fi} \to - l^{\text{cent}}_{fi}$ when $ \phi_0 = 0 \to \pi/3$, and $l^{\text{cent}}_{\lambda} \to - l^{\text{cent}}_{\lambda}$ when $\text{K} \to \text{K}'$, the spectra in Fig. \ref{F3}b at $\phi_0=\pi/3$ can likewise be rationalized in terms of PAM conservation. Therefore, given the ability to perform the identified pre- and post-selected pinwheel state measurements with control over $\mathbf{R}_0$, $\sigma$, and $\phi_0$, it possible to perform energy-, valley, and PAM-resolved inelastic free electron scattering measurements capable of accessing valley phonons at the BZ corners directly.   
}

\textcolor{black}{
As a point of comparison, Fig. \ref{F3}c considers monolayer graphene, where the presence of parity symmetry requires modes $\lambda = 2,3$ to be degenerate at the $\text{K}/\text{K}'$ valleys. Although eigenvectors $\boldsymbol{\xi}_{\lambda=2,3}(\mathbf{k}_{\parallel}={\bf \text{K}}/{\bf \text{K}}')$ can be defined with well-defined parity, parity and PAM operators do not commute. Nevertheless, linear combinations of the parity states can be taken to construct eigenvectors in this degenerate space with well-defined PAM. These particular eigenvectors have the same symmetries as those encountered for hBN, i.e., $\boldsymbol{\xi}_{\lambda=2}$ and $\boldsymbol{\xi}_{\lambda=3}$ exhibit circular motion of carbon atoms on one or the other atomic sites, respectively, while the other atom remains stationary at its equilibrium position. As a consequence, at the $\text{K}$ valleys, these degenerate modes have total PAM $l^{\text{cent}}_{\lambda =2} = +1$ and $l^{\text{cent}}_{\lambda =3} = -1$. Critically, Figs. \ref{F3}b,c show that by comparing the PAM-resolved loss spectra measured at $\mathbf{R}_0 = (0, \pm a/2)$ and $\sigma = \pm1$, it is possible to directly assess non-degeneracy of valley phonon modes with PAM $l_{\lambda}^{\text{cent}} = \pm 1$ introduced in the absence of parity symmetry. Spatial maps of $w^{\phi_0}_{\sigma}(\hbar \omega_{if}, \mathbf{R}_0)$ for hBN and graphene and $\mathbf{R}_0$ varied throughout the real-space unit cell are presented in the SM \cite{SM_chiralphonon}.   
}

Chiral phonons with valley pseudospin degrees of freedom and well-defined PAM underlie a diversity of fundamental condensed matter physics phenomena, material properties, and downstream technologies exploiting unidirectional and lossless phonon propagation in the infrared. However, direct characterization of chiral phonons has remained an outstanding challenge \cite{wang2024chiral} owing to the small linear momentum of light and the absence of obvious polarization degrees of freedom of massive particles. Here we show how pre- and post-selection of pinwheel free electron states during inelastic scattering overcomes these limitations and introduces a new PAM degree of freedom enabling tracking of PAM exchange between probe and sample. By tailoring such pinwheel states to generate 2D periodic arrays of in-plane field vortices with {\AA}-scale polarization textures, we introduce a strategy for direct and selective detection of phonon spin textures of the desired chiral mode symmetries at the $\textrm{K/K}'$ valleys using phase-structured free electrons. Other material systems hosting chiral phonons with various composition, symmetry, and dimensionality could be characterized using generalizations of the proposed inelastic scattering measurements.

\begin{acknowledgments}
We acknowledge Jordan A. Hachtel for valuable discussions and feedback on the strategy proposed. All work was supported by the U.S. Department of Energy (DOE), Office of Science, Office of Basic Energy Sciences (BES), Materials Sciences and Engineering Division under Award No. DOE BES DE-SC0022921. 
\end{acknowledgments}


\bibliography{refs}

\begin{thebibliography}{65}%
\makeatletter
\providecommand \@ifxundefined [1]{%
 \@ifx{#1\undefined}
}%
\providecommand \@ifnum [1]{%
 \ifnum #1\expandafter \@firstoftwo
 \else \expandafter \@secondoftwo
 \fi
}%
\providecommand \@ifx [1]{%
 \ifx #1\expandafter \@firstoftwo
 \else \expandafter \@secondoftwo
 \fi
}%
\providecommand \natexlab [1]{#1}%
\providecommand \enquote  [1]{``#1''}%
\providecommand \bibnamefont  [1]{#1}%
\providecommand \bibfnamefont [1]{#1}%
\providecommand \citenamefont [1]{#1}%
\providecommand \href@noop [0]{\@secondoftwo}%
\providecommand \href [0]{\begingroup \@sanitize@url \@href}%
\providecommand \@href[1]{\@@startlink{#1}\@@href}%
\providecommand \@@href[1]{\endgroup#1\@@endlink}%
\providecommand \@sanitize@url [0]{\catcode `\\12\catcode `\$12\catcode
  `\&12\catcode `\#12\catcode `\^12\catcode `\_12\catcode `\%12\relax}%
\providecommand \@@startlink[1]{}%
\providecommand \@@endlink[0]{}%
\providecommand \url  [0]{\begingroup\@sanitize@url \@url }%
\providecommand \@url [1]{\endgroup\@href {#1}{\urlprefix }}%
\providecommand \urlprefix  [0]{URL }%
\providecommand \Eprint [0]{\href }%
\providecommand \doibase [0]{https://doi.org/}%
\providecommand \selectlanguage [0]{\@gobble}%
\providecommand \bibinfo  [0]{\@secondoftwo}%
\providecommand \bibfield  [0]{\@secondoftwo}%
\providecommand \translation [1]{[#1]}%
\providecommand \BibitemOpen [0]{}%
\providecommand \bibitemStop [0]{}%
\providecommand \bibitemNoStop [0]{.\EOS\space}%
\providecommand \EOS [0]{\spacefactor3000\relax}%
\providecommand \BibitemShut  [1]{\csname bibitem#1\endcsname}%
\let\auto@bib@innerbib\@empty
\bibitem [{\citenamefont {Inomata}\ and\ \citenamefont
  {Kamionkowski}(2019)}]{inomata2019chiral}%
  \BibitemOpen
  \bibfield  {author} {\bibinfo {author} {\bibfnamefont {K.}~\bibnamefont
  {Inomata}}\ and\ \bibinfo {author} {\bibfnamefont {M.}~\bibnamefont
  {Kamionkowski}},\ }\bibfield  {title} {\bibinfo {title} {Chiral photons from
  chiral gravitational waves},\ }\href@noop {} {\bibfield  {journal} {\bibinfo
  {journal} {Phys. Rev. Lett.}\ }\textbf {\bibinfo {volume} {123}},\ \bibinfo
  {pages} {031305} (\bibinfo {year} {2019})}\BibitemShut {NoStop}%
\bibitem [{\citenamefont {Al~Khawaja}\ and\ \citenamefont
  {Stoof}(2001)}]{al2001skyrmions}%
  \BibitemOpen
  \bibfield  {author} {\bibinfo {author} {\bibfnamefont {U.}~\bibnamefont
  {Al~Khawaja}}\ and\ \bibinfo {author} {\bibfnamefont {H.}~\bibnamefont
  {Stoof}},\ }\bibfield  {title} {\bibinfo {title} {Skyrmions in a
  ferromagnetic {B}ose-{E}instein condensate},\ }\href@noop {} {\bibfield
  {journal} {\bibinfo  {journal} {Nature}\ }\textbf {\bibinfo {volume} {411}},\
  \bibinfo {pages} {918} (\bibinfo {year} {2001})}\BibitemShut {NoStop}%
\bibitem [{\citenamefont {Yu}\ \emph {et~al.}(2018)\citenamefont {Yu},
  \citenamefont {Koshibae}, \citenamefont {Tokunaga}, \citenamefont {Shibata},
  \citenamefont {Taguchi}, \citenamefont {Nagaosa},\ and\ \citenamefont
  {Tokura}}]{yu2018transformation}%
  \BibitemOpen
  \bibfield  {author} {\bibinfo {author} {\bibfnamefont {X.}~\bibnamefont
  {Yu}}, \bibinfo {author} {\bibfnamefont {W.}~\bibnamefont {Koshibae}},
  \bibinfo {author} {\bibfnamefont {Y.}~\bibnamefont {Tokunaga}}, \bibinfo
  {author} {\bibfnamefont {K.}~\bibnamefont {Shibata}}, \bibinfo {author}
  {\bibfnamefont {Y.}~\bibnamefont {Taguchi}}, \bibinfo {author} {\bibfnamefont
  {N.}~\bibnamefont {Nagaosa}},\ and\ \bibinfo {author} {\bibfnamefont
  {Y.}~\bibnamefont {Tokura}},\ }\bibfield  {title} {\bibinfo {title}
  {Transformation between meron and skyrmion topological spin textures in a
  chiral magnet},\ }\href@noop {} {\bibfield  {journal} {\bibinfo  {journal}
  {Nature}\ }\textbf {\bibinfo {volume} {564}},\ \bibinfo {pages} {95}
  (\bibinfo {year} {2018})}\BibitemShut {NoStop}%
\bibitem [{\citenamefont {Gao}\ \emph {et~al.}(2020)\citenamefont {Gao},
  \citenamefont {Speirits}, \citenamefont {Castellucci}, \citenamefont
  {Franke-Arnold}, \citenamefont {Barnett},\ and\ \citenamefont
  {G{\"o}tte}}]{gao2020paraxial}%
  \BibitemOpen
  \bibfield  {author} {\bibinfo {author} {\bibfnamefont {S.}~\bibnamefont
  {Gao}}, \bibinfo {author} {\bibfnamefont {F.~C.}\ \bibnamefont {Speirits}},
  \bibinfo {author} {\bibfnamefont {F.}~\bibnamefont {Castellucci}}, \bibinfo
  {author} {\bibfnamefont {S.}~\bibnamefont {Franke-Arnold}}, \bibinfo {author}
  {\bibfnamefont {S.~M.}\ \bibnamefont {Barnett}},\ and\ \bibinfo {author}
  {\bibfnamefont {J.~B.}\ \bibnamefont {G{\"o}tte}},\ }\bibfield  {title}
  {\bibinfo {title} {Paraxial skyrmionic beams},\ }\href@noop {} {\bibfield
  {journal} {\bibinfo  {journal} {Phys. Rev. A}\ }\textbf {\bibinfo {volume}
  {102}},\ \bibinfo {pages} {053513} (\bibinfo {year} {2020})}\BibitemShut
  {NoStop}%
\bibitem [{\citenamefont {Shen}\ \emph {et~al.}(2021)\citenamefont {Shen},
  \citenamefont {Hou}, \citenamefont {Papasimakis},\ and\ \citenamefont
  {Zheludev}}]{shen2021supertoroidal}%
  \BibitemOpen
  \bibfield  {author} {\bibinfo {author} {\bibfnamefont {Y.}~\bibnamefont
  {Shen}}, \bibinfo {author} {\bibfnamefont {Y.}~\bibnamefont {Hou}}, \bibinfo
  {author} {\bibfnamefont {N.}~\bibnamefont {Papasimakis}},\ and\ \bibinfo
  {author} {\bibfnamefont {N.~I.}\ \bibnamefont {Zheludev}},\ }\bibfield
  {title} {\bibinfo {title} {Supertoroidal light pulses as electromagnetic
  skyrmions propagating in free space},\ }\href@noop {} {\bibfield  {journal}
  {\bibinfo  {journal} {Nat. Commun.}\ }\textbf {\bibinfo {volume} {12}},\
  \bibinfo {pages} {5891} (\bibinfo {year} {2021})}\BibitemShut {NoStop}%
\bibitem [{\citenamefont {Tsesses}\ \emph {et~al.}(2018)\citenamefont
  {Tsesses}, \citenamefont {Ostrovsky}, \citenamefont {Cohen}, \citenamefont
  {Gjonaj}, \citenamefont {Lindner},\ and\ \citenamefont
  {Bartal}}]{tsesses2018optical}%
  \BibitemOpen
  \bibfield  {author} {\bibinfo {author} {\bibfnamefont {S.}~\bibnamefont
  {Tsesses}}, \bibinfo {author} {\bibfnamefont {E.}~\bibnamefont {Ostrovsky}},
  \bibinfo {author} {\bibfnamefont {K.}~\bibnamefont {Cohen}}, \bibinfo
  {author} {\bibfnamefont {B.}~\bibnamefont {Gjonaj}}, \bibinfo {author}
  {\bibfnamefont {N.}~\bibnamefont {Lindner}},\ and\ \bibinfo {author}
  {\bibfnamefont {G.}~\bibnamefont {Bartal}},\ }\bibfield  {title} {\bibinfo
  {title} {Optical skyrmion lattice in evanescent electromagnetic fields},\
  }\href@noop {} {\bibfield  {journal} {\bibinfo  {journal} {Science}\ }\textbf
  {\bibinfo {volume} {361}},\ \bibinfo {pages} {993} (\bibinfo {year}
  {2018})}\BibitemShut {NoStop}%
\bibitem [{\citenamefont {Du}\ \emph {et~al.}(2019)\citenamefont {Du},
  \citenamefont {Yang}, \citenamefont {Zayats},\ and\ \citenamefont
  {Yuan}}]{du2019deep}%
  \BibitemOpen
  \bibfield  {author} {\bibinfo {author} {\bibfnamefont {L.}~\bibnamefont
  {Du}}, \bibinfo {author} {\bibfnamefont {A.}~\bibnamefont {Yang}}, \bibinfo
  {author} {\bibfnamefont {A.~V.}\ \bibnamefont {Zayats}},\ and\ \bibinfo
  {author} {\bibfnamefont {X.}~\bibnamefont {Yuan}},\ }\bibfield  {title}
  {\bibinfo {title} {Deep-subwavelength features of photonic skyrmions in a
  confined electromagnetic field with orbital angular momentum},\ }\href@noop
  {} {\bibfield  {journal} {\bibinfo  {journal} {Nat. Phys.}\ }\textbf
  {\bibinfo {volume} {15}},\ \bibinfo {pages} {650} (\bibinfo {year}
  {2019})}\BibitemShut {NoStop}%
\bibitem [{\citenamefont {Dai}\ \emph {et~al.}(2020)\citenamefont {Dai},
  \citenamefont {Zhou}, \citenamefont {Ghosh}, \citenamefont {Mong},
  \citenamefont {Kubo}, \citenamefont {Huang},\ and\ \citenamefont
  {Petek}}]{dai2020plasmonic}%
  \BibitemOpen
  \bibfield  {author} {\bibinfo {author} {\bibfnamefont {Y.}~\bibnamefont
  {Dai}}, \bibinfo {author} {\bibfnamefont {Z.}~\bibnamefont {Zhou}}, \bibinfo
  {author} {\bibfnamefont {A.}~\bibnamefont {Ghosh}}, \bibinfo {author}
  {\bibfnamefont {R.~S.}\ \bibnamefont {Mong}}, \bibinfo {author}
  {\bibfnamefont {A.}~\bibnamefont {Kubo}}, \bibinfo {author} {\bibfnamefont
  {C.-B.}\ \bibnamefont {Huang}},\ and\ \bibinfo {author} {\bibfnamefont
  {H.}~\bibnamefont {Petek}},\ }\bibfield  {title} {\bibinfo {title} {Plasmonic
  topological quasiparticle on the nanometre and femtosecond scales},\
  }\href@noop {} {\bibfield  {journal} {\bibinfo  {journal} {Nature}\ }\textbf
  {\bibinfo {volume} {588}},\ \bibinfo {pages} {616} (\bibinfo {year}
  {2020})}\BibitemShut {NoStop}%
\bibitem [{\citenamefont {Davis}\ \emph {et~al.}(2020)\citenamefont {Davis},
  \citenamefont {Janoschka}, \citenamefont {Dreher}, \citenamefont {Frank},
  \citenamefont {Meyer~zu Heringdorf},\ and\ \citenamefont
  {Giessen}}]{davis2020ultrafast}%
  \BibitemOpen
  \bibfield  {author} {\bibinfo {author} {\bibfnamefont {T.~J.}\ \bibnamefont
  {Davis}}, \bibinfo {author} {\bibfnamefont {D.}~\bibnamefont {Janoschka}},
  \bibinfo {author} {\bibfnamefont {P.}~\bibnamefont {Dreher}}, \bibinfo
  {author} {\bibfnamefont {B.}~\bibnamefont {Frank}}, \bibinfo {author}
  {\bibfnamefont {F.-J.}\ \bibnamefont {Meyer~zu Heringdorf}},\ and\ \bibinfo
  {author} {\bibfnamefont {H.}~\bibnamefont {Giessen}},\ }\bibfield  {title}
  {\bibinfo {title} {Ultrafast vector imaging of plasmonic skyrmion dynamics
  with deep subwavelength resolution},\ }\href@noop {} {\bibfield  {journal}
  {\bibinfo  {journal} {Science}\ }\textbf {\bibinfo {volume} {368}},\ \bibinfo
  {pages} {eaba6415} (\bibinfo {year} {2020})}\BibitemShut {NoStop}%
\bibitem [{\citenamefont {Ghosh}\ \emph {et~al.}(2023)\citenamefont {Ghosh},
  \citenamefont {Yang}, \citenamefont {Dai},\ and\ \citenamefont
  {Petek}}]{ghosh2023spin}%
  \BibitemOpen
  \bibfield  {author} {\bibinfo {author} {\bibfnamefont {A.}~\bibnamefont
  {Ghosh}}, \bibinfo {author} {\bibfnamefont {S.}~\bibnamefont {Yang}},
  \bibinfo {author} {\bibfnamefont {Y.}~\bibnamefont {Dai}},\ and\ \bibinfo
  {author} {\bibfnamefont {H.}~\bibnamefont {Petek}},\ }\bibfield  {title}
  {\bibinfo {title} {The spin texture topology of polygonal plasmon fields},\
  }\href@noop {} {\bibfield  {journal} {\bibinfo  {journal} {ACS Photonics}\
  }\textbf {\bibinfo {volume} {10}},\ \bibinfo {pages} {13} (\bibinfo {year}
  {2023})}\BibitemShut {NoStop}%
\bibitem [{\citenamefont {Chen}\ \emph {et~al.}(2015)\citenamefont {Chen},
  \citenamefont {Zheng}, \citenamefont {Fuhrer},\ and\ \citenamefont
  {Yan}}]{chen2015helicity}%
  \BibitemOpen
  \bibfield  {author} {\bibinfo {author} {\bibfnamefont {S.-Y.}\ \bibnamefont
  {Chen}}, \bibinfo {author} {\bibfnamefont {C.}~\bibnamefont {Zheng}},
  \bibinfo {author} {\bibfnamefont {M.~S.}\ \bibnamefont {Fuhrer}},\ and\
  \bibinfo {author} {\bibfnamefont {J.}~\bibnamefont {Yan}},\ }\bibfield
  {title} {\bibinfo {title} {Helicity-resolved {R}aman scattering of
  {M}o{S}$_2$, {M}o{S}e$_2$, {W}{S}$_2$, and {W}{S}e$_2$ atomic layers},\
  }\href@noop {} {\bibfield  {journal} {\bibinfo  {journal} {Nano Lett.}\
  }\textbf {\bibinfo {volume} {15}},\ \bibinfo {pages} {2526} (\bibinfo {year}
  {2015})}\BibitemShut {NoStop}%
\bibitem [{\citenamefont {Zhang}\ and\ \citenamefont
  {Niu}(2015)}]{zhang2015chiral}%
  \BibitemOpen
  \bibfield  {author} {\bibinfo {author} {\bibfnamefont {L.}~\bibnamefont
  {Zhang}}\ and\ \bibinfo {author} {\bibfnamefont {Q.}~\bibnamefont {Niu}},\
  }\bibfield  {title} {\bibinfo {title} {Chiral phonons at high-symmetry points
  in monolayer hexagonal lattices},\ }\href@noop {} {\bibfield  {journal}
  {\bibinfo  {journal} {Phys. Rev. Lett.}\ }\textbf {\bibinfo {volume} {115}},\
  \bibinfo {pages} {115502} (\bibinfo {year} {2015})}\BibitemShut {NoStop}%
\bibitem [{\citenamefont {Zhu}\ \emph {et~al.}(2018)\citenamefont {Zhu},
  \citenamefont {Yi}, \citenamefont {Li}, \citenamefont {Xiao}, \citenamefont
  {Zhang}, \citenamefont {Yang}, \citenamefont {Kaindl}, \citenamefont {Li},
  \citenamefont {Wang},\ and\ \citenamefont {Zhang}}]{zhu2018observation}%
  \BibitemOpen
  \bibfield  {author} {\bibinfo {author} {\bibfnamefont {H.}~\bibnamefont
  {Zhu}}, \bibinfo {author} {\bibfnamefont {J.}~\bibnamefont {Yi}}, \bibinfo
  {author} {\bibfnamefont {M.-Y.}\ \bibnamefont {Li}}, \bibinfo {author}
  {\bibfnamefont {J.}~\bibnamefont {Xiao}}, \bibinfo {author} {\bibfnamefont
  {L.}~\bibnamefont {Zhang}}, \bibinfo {author} {\bibfnamefont {C.-W.}\
  \bibnamefont {Yang}}, \bibinfo {author} {\bibfnamefont {R.~A.}\ \bibnamefont
  {Kaindl}}, \bibinfo {author} {\bibfnamefont {L.-J.}\ \bibnamefont {Li}},
  \bibinfo {author} {\bibfnamefont {Y.}~\bibnamefont {Wang}},\ and\ \bibinfo
  {author} {\bibfnamefont {X.}~\bibnamefont {Zhang}},\ }\bibfield  {title}
  {\bibinfo {title} {Observation of chiral phonons},\ }\href@noop {} {\bibfield
   {journal} {\bibinfo  {journal} {Science}\ }\textbf {\bibinfo {volume}
  {359}},\ \bibinfo {pages} {579} (\bibinfo {year} {2018})}\BibitemShut
  {NoStop}%
\bibitem [{\citenamefont {Chen}\ \emph {et~al.}(2018)\citenamefont {Chen},
  \citenamefont {Zhang}, \citenamefont {Niu},\ and\ \citenamefont
  {Zhang}}]{chen2018chiral}%
  \BibitemOpen
  \bibfield  {author} {\bibinfo {author} {\bibfnamefont {H.}~\bibnamefont
  {Chen}}, \bibinfo {author} {\bibfnamefont {W.}~\bibnamefont {Zhang}},
  \bibinfo {author} {\bibfnamefont {Q.}~\bibnamefont {Niu}},\ and\ \bibinfo
  {author} {\bibfnamefont {L.}~\bibnamefont {Zhang}},\ }\bibfield  {title}
  {\bibinfo {title} {Chiral phonons in two-dimensional materials},\ }\href@noop
  {} {\bibfield  {journal} {\bibinfo  {journal} {2D Mater.}\ }\textbf {\bibinfo
  {volume} {6}},\ \bibinfo {pages} {012002} (\bibinfo {year}
  {2018})}\BibitemShut {NoStop}%
\bibitem [{\citenamefont {Chen}\ \emph {et~al.}(2019)\citenamefont {Chen},
  \citenamefont {Wu}, \citenamefont {Yang}, \citenamefont {Li},\ and\
  \citenamefont {Zhang}}]{chen2019chiral}%
  \BibitemOpen
  \bibfield  {author} {\bibinfo {author} {\bibfnamefont {H.}~\bibnamefont
  {Chen}}, \bibinfo {author} {\bibfnamefont {W.}~\bibnamefont {Wu}}, \bibinfo
  {author} {\bibfnamefont {S.~A.}\ \bibnamefont {Yang}}, \bibinfo {author}
  {\bibfnamefont {X.}~\bibnamefont {Li}},\ and\ \bibinfo {author}
  {\bibfnamefont {L.}~\bibnamefont {Zhang}},\ }\bibfield  {title} {\bibinfo
  {title} {Chiral phonons in kagome lattices},\ }\href@noop {} {\bibfield
  {journal} {\bibinfo  {journal} {Phys. Rev. B}\ }\textbf {\bibinfo {volume}
  {100}},\ \bibinfo {pages} {094303} (\bibinfo {year} {2019})}\BibitemShut
  {NoStop}%
\bibitem [{\citenamefont {Suri}\ \emph {et~al.}(2021)\citenamefont {Suri},
  \citenamefont {Wang}, \citenamefont {Zhang},\ and\ \citenamefont
  {Xiao}}]{suri2021chiral}%
  \BibitemOpen
  \bibfield  {author} {\bibinfo {author} {\bibfnamefont {N.}~\bibnamefont
  {Suri}}, \bibinfo {author} {\bibfnamefont {C.}~\bibnamefont {Wang}}, \bibinfo
  {author} {\bibfnamefont {Y.}~\bibnamefont {Zhang}},\ and\ \bibinfo {author}
  {\bibfnamefont {D.}~\bibnamefont {Xiao}},\ }\bibfield  {title} {\bibinfo
  {title} {Chiral phonons in moir{\'e} superlattices},\ }\href@noop {}
  {\bibfield  {journal} {\bibinfo  {journal} {Nano Lett.}\ }\textbf {\bibinfo
  {volume} {21}},\ \bibinfo {pages} {10026} (\bibinfo {year}
  {2021})}\BibitemShut {NoStop}%
\bibitem [{\citenamefont {Wang}\ \emph {et~al.}(2022)\citenamefont {Wang},
  \citenamefont {Li}, \citenamefont {Zhu}, \citenamefont {Chen}, \citenamefont
  {Wu}, \citenamefont {Gao}, \citenamefont {Zhang},\ and\ \citenamefont
  {Yang}}]{wang2022chiral}%
  \BibitemOpen
  \bibfield  {author} {\bibinfo {author} {\bibfnamefont {Q.}~\bibnamefont
  {Wang}}, \bibinfo {author} {\bibfnamefont {S.}~\bibnamefont {Li}}, \bibinfo
  {author} {\bibfnamefont {J.}~\bibnamefont {Zhu}}, \bibinfo {author}
  {\bibfnamefont {H.}~\bibnamefont {Chen}}, \bibinfo {author} {\bibfnamefont
  {W.}~\bibnamefont {Wu}}, \bibinfo {author} {\bibfnamefont {W.}~\bibnamefont
  {Gao}}, \bibinfo {author} {\bibfnamefont {L.}~\bibnamefont {Zhang}},\ and\
  \bibinfo {author} {\bibfnamefont {S.~A.}\ \bibnamefont {Yang}},\ }\bibfield
  {title} {\bibinfo {title} {Chiral phonons in lattices with {C}$_4$
  symmetry},\ }\href@noop {} {\bibfield  {journal} {\bibinfo  {journal} {Phys.
  Rev. B}\ }\textbf {\bibinfo {volume} {105}},\ \bibinfo {pages} {104301}
  (\bibinfo {year} {2022})}\BibitemShut {NoStop}%
\bibitem [{\citenamefont {Zhang}\ and\ \citenamefont
  {Murakami}(2022)}]{zhang2022chiral}%
  \BibitemOpen
  \bibfield  {author} {\bibinfo {author} {\bibfnamefont {T.}~\bibnamefont
  {Zhang}}\ and\ \bibinfo {author} {\bibfnamefont {S.}~\bibnamefont
  {Murakami}},\ }\bibfield  {title} {\bibinfo {title} {Chiral phonons and
  pseudoangular momentum in nonsymmorphic systems},\ }\href@noop {} {\bibfield
  {journal} {\bibinfo  {journal} {Phys. Rev. Research}\ }\textbf {\bibinfo
  {volume} {4}},\ \bibinfo {pages} {L012024} (\bibinfo {year}
  {2022})}\BibitemShut {NoStop}%
\bibitem [{\citenamefont {Fransson}(2023)}]{fransson2023chiral}%
  \BibitemOpen
  \bibfield  {author} {\bibinfo {author} {\bibfnamefont {J.}~\bibnamefont
  {Fransson}},\ }\bibfield  {title} {\bibinfo {title} {Chiral phonon induced
  spin polarization},\ }\href@noop {} {\bibfield  {journal} {\bibinfo
  {journal} {Phys. Rev. Research}\ }\textbf {\bibinfo {volume} {5}},\ \bibinfo
  {pages} {L022039} (\bibinfo {year} {2023})}\BibitemShut {NoStop}%
\bibitem [{\citenamefont {Chen}\ \emph {et~al.}(2021)\citenamefont {Chen},
  \citenamefont {Wu}, \citenamefont {Zhu}, \citenamefont {Yang},\ and\
  \citenamefont {Zhang}}]{chen2021propagating}%
  \BibitemOpen
  \bibfield  {author} {\bibinfo {author} {\bibfnamefont {H.}~\bibnamefont
  {Chen}}, \bibinfo {author} {\bibfnamefont {W.}~\bibnamefont {Wu}}, \bibinfo
  {author} {\bibfnamefont {J.}~\bibnamefont {Zhu}}, \bibinfo {author}
  {\bibfnamefont {S.~A.}\ \bibnamefont {Yang}},\ and\ \bibinfo {author}
  {\bibfnamefont {L.}~\bibnamefont {Zhang}},\ }\bibfield  {title} {\bibinfo
  {title} {Propagating chiral phonons in three-dimensional materials},\
  }\href@noop {} {\bibfield  {journal} {\bibinfo  {journal} {Nano Lett.}\
  }\textbf {\bibinfo {volume} {21}},\ \bibinfo {pages} {3060} (\bibinfo {year}
  {2021})}\BibitemShut {NoStop}%
\bibitem [{\citenamefont {Ishito}\ \emph {et~al.}(2023)\citenamefont {Ishito},
  \citenamefont {Mao}, \citenamefont {Kousaka}, \citenamefont {Togawa},
  \citenamefont {Iwasaki}, \citenamefont {Zhang}, \citenamefont {Murakami},
  \citenamefont {Kishine},\ and\ \citenamefont {Satoh}}]{ishito2023truly}%
  \BibitemOpen
  \bibfield  {author} {\bibinfo {author} {\bibfnamefont {K.}~\bibnamefont
  {Ishito}}, \bibinfo {author} {\bibfnamefont {H.}~\bibnamefont {Mao}},
  \bibinfo {author} {\bibfnamefont {Y.}~\bibnamefont {Kousaka}}, \bibinfo
  {author} {\bibfnamefont {Y.}~\bibnamefont {Togawa}}, \bibinfo {author}
  {\bibfnamefont {S.}~\bibnamefont {Iwasaki}}, \bibinfo {author} {\bibfnamefont
  {T.}~\bibnamefont {Zhang}}, \bibinfo {author} {\bibfnamefont
  {S.}~\bibnamefont {Murakami}}, \bibinfo {author} {\bibfnamefont {J.-i.}\
  \bibnamefont {Kishine}},\ and\ \bibinfo {author} {\bibfnamefont
  {T.}~\bibnamefont {Satoh}},\ }\bibfield  {title} {\bibinfo {title} {Truly
  chiral phonons in $\alpha$-{H}g{S}},\ }\href@noop {} {\bibfield  {journal}
  {\bibinfo  {journal} {Nat. Phys.}\ }\textbf {\bibinfo {volume} {19}},\
  \bibinfo {pages} {35} (\bibinfo {year} {2023})}\BibitemShut {NoStop}%
\bibitem [{\citenamefont {Ueda}\ \emph {et~al.}(2023)\citenamefont {Ueda},
  \citenamefont {Garc{\'\i}a-Fern{\'a}ndez}, \citenamefont {Agrestini},
  \citenamefont {Romao}, \citenamefont {van~den Brink}, \citenamefont
  {Spaldin}, \citenamefont {Zhou},\ and\ \citenamefont
  {Staub}}]{ueda2023chiral}%
  \BibitemOpen
  \bibfield  {author} {\bibinfo {author} {\bibfnamefont {H.}~\bibnamefont
  {Ueda}}, \bibinfo {author} {\bibfnamefont {M.}~\bibnamefont
  {Garc{\'\i}a-Fern{\'a}ndez}}, \bibinfo {author} {\bibfnamefont
  {S.}~\bibnamefont {Agrestini}}, \bibinfo {author} {\bibfnamefont {C.~P.}\
  \bibnamefont {Romao}}, \bibinfo {author} {\bibfnamefont {J.}~\bibnamefont
  {van~den Brink}}, \bibinfo {author} {\bibfnamefont {N.~A.}\ \bibnamefont
  {Spaldin}}, \bibinfo {author} {\bibfnamefont {K.-J.}\ \bibnamefont {Zhou}},\
  and\ \bibinfo {author} {\bibfnamefont {U.}~\bibnamefont {Staub}},\ }\bibfield
   {title} {\bibinfo {title} {Chiral phonons in quartz probed by x-rays},\
  }\href@noop {} {\bibfield  {journal} {\bibinfo  {journal} {Nature}\ ,\
  \bibinfo {pages} {1}} (\bibinfo {year} {2023})}\BibitemShut {NoStop}%
\bibitem [{\citenamefont {Grissonnanche}\ \emph {et~al.}(2020)\citenamefont
  {Grissonnanche}, \citenamefont {Th{\'e}riault}, \citenamefont {Gourgout},
  \citenamefont {Boulanger}, \citenamefont {Lefran{\c{c}}ois}, \citenamefont
  {Ataei}, \citenamefont {Lalibert{\'e}}, \citenamefont {Dion}, \citenamefont
  {Zhou}, \citenamefont {Pyon}, \citenamefont {Takayama}, \citenamefont
  {Takagi}, \citenamefont {Doiron-Leyraud},\ and\ \citenamefont
  {Taillefer}}]{grissonnanche2020chiral}%
  \BibitemOpen
  \bibfield  {author} {\bibinfo {author} {\bibfnamefont {G.}~\bibnamefont
  {Grissonnanche}}, \bibinfo {author} {\bibfnamefont {S.}~\bibnamefont
  {Th{\'e}riault}}, \bibinfo {author} {\bibfnamefont {A.}~\bibnamefont
  {Gourgout}}, \bibinfo {author} {\bibfnamefont {M.-E.}\ \bibnamefont
  {Boulanger}}, \bibinfo {author} {\bibfnamefont {E.}~\bibnamefont
  {Lefran{\c{c}}ois}}, \bibinfo {author} {\bibfnamefont {A.}~\bibnamefont
  {Ataei}}, \bibinfo {author} {\bibfnamefont {F.}~\bibnamefont
  {Lalibert{\'e}}}, \bibinfo {author} {\bibfnamefont {M.}~\bibnamefont {Dion}},
  \bibinfo {author} {\bibfnamefont {J.-S.}\ \bibnamefont {Zhou}}, \bibinfo
  {author} {\bibfnamefont {S.}~\bibnamefont {Pyon}}, \bibinfo {author}
  {\bibfnamefont {T.}~\bibnamefont {Takayama}}, \bibinfo {author}
  {\bibfnamefont {H.}~\bibnamefont {Takagi}}, \bibinfo {author} {\bibfnamefont
  {N.}~\bibnamefont {Doiron-Leyraud}},\ and\ \bibinfo {author} {\bibfnamefont
  {L.}~\bibnamefont {Taillefer}},\ }\bibfield  {title} {\bibinfo {title}
  {Chiral phonons in the pseudogap phase of cuprates},\ }\href@noop {}
  {\bibfield  {journal} {\bibinfo  {journal} {Nat. Phys.}\ }\textbf {\bibinfo
  {volume} {16}},\ \bibinfo {pages} {1108} (\bibinfo {year}
  {2020})}\BibitemShut {NoStop}%
\bibitem [{\citenamefont {Kim}\ \emph {et~al.}(2023)\citenamefont {Kim},
  \citenamefont {Vetter}, \citenamefont {Yan}, \citenamefont {Yang},
  \citenamefont {Wang}, \citenamefont {Sun}, \citenamefont {Yang},
  \citenamefont {Comstock}, \citenamefont {Li}, \citenamefont {Zhou},
  \citenamefont {Zhang}, \citenamefont {You}, \citenamefont {Sun},\ and\
  \citenamefont {Liu}}]{kim2023chiral}%
  \BibitemOpen
  \bibfield  {author} {\bibinfo {author} {\bibfnamefont {K.}~\bibnamefont
  {Kim}}, \bibinfo {author} {\bibfnamefont {E.}~\bibnamefont {Vetter}},
  \bibinfo {author} {\bibfnamefont {L.}~\bibnamefont {Yan}}, \bibinfo {author}
  {\bibfnamefont {C.}~\bibnamefont {Yang}}, \bibinfo {author} {\bibfnamefont
  {Z.}~\bibnamefont {Wang}}, \bibinfo {author} {\bibfnamefont {R.}~\bibnamefont
  {Sun}}, \bibinfo {author} {\bibfnamefont {Y.}~\bibnamefont {Yang}}, \bibinfo
  {author} {\bibfnamefont {A.~H.}\ \bibnamefont {Comstock}}, \bibinfo {author}
  {\bibfnamefont {X.}~\bibnamefont {Li}}, \bibinfo {author} {\bibfnamefont
  {J.}~\bibnamefont {Zhou}}, \bibinfo {author} {\bibfnamefont {L.}~\bibnamefont
  {Zhang}}, \bibinfo {author} {\bibfnamefont {W.}~\bibnamefont {You}}, \bibinfo
  {author} {\bibfnamefont {D.}~\bibnamefont {Sun}},\ and\ \bibinfo {author}
  {\bibfnamefont {J.}~\bibnamefont {Liu}},\ }\bibfield  {title} {\bibinfo
  {title} {Chiral-phonon-activated spin {S}eebeck effect},\ }\href@noop {}
  {\bibfield  {journal} {\bibinfo  {journal} {Nat. Mater.}\ }\textbf {\bibinfo
  {volume} {22}},\ \bibinfo {pages} {322} (\bibinfo {year} {2023})}\BibitemShut
  {NoStop}%
\bibitem [{\citenamefont {Jeong}\ \emph {et~al.}(2022)\citenamefont {Jeong},
  \citenamefont {Kim}, \citenamefont {Seo}, \citenamefont {Park}, \citenamefont
  {Jeong}, \citenamefont {Kim}, \citenamefont {Lauter}, \citenamefont {Egami},
  \citenamefont {Han},\ and\ \citenamefont {Choi}}]{jeong2022unconventional}%
  \BibitemOpen
  \bibfield  {author} {\bibinfo {author} {\bibfnamefont {S.~G.}\ \bibnamefont
  {Jeong}}, \bibinfo {author} {\bibfnamefont {J.}~\bibnamefont {Kim}}, \bibinfo
  {author} {\bibfnamefont {A.}~\bibnamefont {Seo}}, \bibinfo {author}
  {\bibfnamefont {S.}~\bibnamefont {Park}}, \bibinfo {author} {\bibfnamefont
  {H.~Y.}\ \bibnamefont {Jeong}}, \bibinfo {author} {\bibfnamefont {Y.-M.}\
  \bibnamefont {Kim}}, \bibinfo {author} {\bibfnamefont {V.}~\bibnamefont
  {Lauter}}, \bibinfo {author} {\bibfnamefont {T.}~\bibnamefont {Egami}},
  \bibinfo {author} {\bibfnamefont {J.~H.}\ \bibnamefont {Han}},\ and\ \bibinfo
  {author} {\bibfnamefont {W.~S.}\ \bibnamefont {Choi}},\ }\bibfield  {title}
  {\bibinfo {title} {Unconventional interlayer exchange coupling via chiral
  phonons in synthetic magnetic oxide heterostructures},\ }\href@noop {}
  {\bibfield  {journal} {\bibinfo  {journal} {Sci. Adv.}\ }\textbf {\bibinfo
  {volume} {8}},\ \bibinfo {pages} {eabm4005} (\bibinfo {year}
  {2022})}\BibitemShut {NoStop}%
\bibitem [{\citenamefont {Chen}\ \emph {et~al.}(2022)\citenamefont {Chen},
  \citenamefont {Wu}, \citenamefont {Zhu}, \citenamefont {Yang}, \citenamefont
  {Gong}, \citenamefont {Gao}, \citenamefont {Yang},\ and\ \citenamefont
  {Zhang}}]{chen2022chiral}%
  \BibitemOpen
  \bibfield  {author} {\bibinfo {author} {\bibfnamefont {H.}~\bibnamefont
  {Chen}}, \bibinfo {author} {\bibfnamefont {W.}~\bibnamefont {Wu}}, \bibinfo
  {author} {\bibfnamefont {J.}~\bibnamefont {Zhu}}, \bibinfo {author}
  {\bibfnamefont {Z.}~\bibnamefont {Yang}}, \bibinfo {author} {\bibfnamefont
  {W.}~\bibnamefont {Gong}}, \bibinfo {author} {\bibfnamefont {W.}~\bibnamefont
  {Gao}}, \bibinfo {author} {\bibfnamefont {S.~A.}\ \bibnamefont {Yang}},\ and\
  \bibinfo {author} {\bibfnamefont {L.}~\bibnamefont {Zhang}},\ }\bibfield
  {title} {\bibinfo {title} {Chiral phonon diode effect in chiral crystals},\
  }\href@noop {} {\bibfield  {journal} {\bibinfo  {journal} {Nano Lett.}\
  }\textbf {\bibinfo {volume} {22}},\ \bibinfo {pages} {1688} (\bibinfo {year}
  {2022})}\BibitemShut {NoStop}%
\bibitem [{\citenamefont {Wang}\ \emph {et~al.}(2024)\citenamefont {Wang},
  \citenamefont {Sun}, \citenamefont {Li},\ and\ \citenamefont
  {Zhang}}]{wang2024chiral}%
  \BibitemOpen
  \bibfield  {author} {\bibinfo {author} {\bibfnamefont {T.}~\bibnamefont
  {Wang}}, \bibinfo {author} {\bibfnamefont {H.}~\bibnamefont {Sun}}, \bibinfo
  {author} {\bibfnamefont {X.}~\bibnamefont {Li}},\ and\ \bibinfo {author}
  {\bibfnamefont {L.}~\bibnamefont {Zhang}},\ }\bibfield  {title} {\bibinfo
  {title} {Chiral phonons: Prediction, verification, and application},\
  }\href@noop {} {\bibfield  {journal} {\bibinfo  {journal} {Nano Lett.}\
  }\textbf {\bibinfo {volume} {24}},\ \bibinfo {pages} {4311} (\bibinfo {year}
  {2024})}\BibitemShut {NoStop}%
\bibitem [{\citenamefont {Krivanek}\ \emph {et~al.}(2014)\citenamefont
  {Krivanek}, \citenamefont {Lovejoy}, \citenamefont {Dellby}, \citenamefont
  {Aoki}, \citenamefont {Carpenter}, \citenamefont {Rez}, \citenamefont
  {Soignard}, \citenamefont {Zhu}, \citenamefont {Batson}, \citenamefont
  {Lagos}, \citenamefont {Egerton},\ and\ \citenamefont
  {Crozier}}]{krivanek2014vibrational}%
  \BibitemOpen
  \bibfield  {author} {\bibinfo {author} {\bibfnamefont {O.~L.}\ \bibnamefont
  {Krivanek}}, \bibinfo {author} {\bibfnamefont {T.~C.}\ \bibnamefont
  {Lovejoy}}, \bibinfo {author} {\bibfnamefont {N.}~\bibnamefont {Dellby}},
  \bibinfo {author} {\bibfnamefont {T.}~\bibnamefont {Aoki}}, \bibinfo {author}
  {\bibfnamefont {R.}~\bibnamefont {Carpenter}}, \bibinfo {author}
  {\bibfnamefont {P.}~\bibnamefont {Rez}}, \bibinfo {author} {\bibfnamefont
  {E.}~\bibnamefont {Soignard}}, \bibinfo {author} {\bibfnamefont
  {J.}~\bibnamefont {Zhu}}, \bibinfo {author} {\bibfnamefont {P.~E.}\
  \bibnamefont {Batson}}, \bibinfo {author} {\bibfnamefont {M.~J.}\
  \bibnamefont {Lagos}}, \bibinfo {author} {\bibfnamefont {R.~F.}\ \bibnamefont
  {Egerton}},\ and\ \bibinfo {author} {\bibfnamefont {P.~A.}\ \bibnamefont
  {Crozier}},\ }\bibfield  {title} {\bibinfo {title} {Vibrational spectroscopy
  in the electron microscope},\ }\href@noop {} {\bibfield  {journal} {\bibinfo
  {journal} {Nature}\ }\textbf {\bibinfo {volume} {514}},\ \bibinfo {pages}
  {209} (\bibinfo {year} {2014})}\BibitemShut {NoStop}%
\bibitem [{\citenamefont {Dwyer}\ \emph {et~al.}(2016)\citenamefont {Dwyer},
  \citenamefont {Aoki}, \citenamefont {Rez}, \citenamefont {Chang},
  \citenamefont {Lovejoy},\ and\ \citenamefont {Krivanek}}]{dwyer2016electron}%
  \BibitemOpen
  \bibfield  {author} {\bibinfo {author} {\bibfnamefont {C.}~\bibnamefont
  {Dwyer}}, \bibinfo {author} {\bibfnamefont {T.}~\bibnamefont {Aoki}},
  \bibinfo {author} {\bibfnamefont {P.}~\bibnamefont {Rez}}, \bibinfo {author}
  {\bibfnamefont {S.}~\bibnamefont {Chang}}, \bibinfo {author} {\bibfnamefont
  {T.}~\bibnamefont {Lovejoy}},\ and\ \bibinfo {author} {\bibfnamefont
  {O.}~\bibnamefont {Krivanek}},\ }\bibfield  {title} {\bibinfo {title}
  {Electron-beam mapping of vibrational modes with nanometer spatial
  resolution},\ }\href@noop {} {\bibfield  {journal} {\bibinfo  {journal}
  {Phys. Rev. Lett.}\ }\textbf {\bibinfo {volume} {117}},\ \bibinfo {pages}
  {256101} (\bibinfo {year} {2016})}\BibitemShut {NoStop}%
\bibitem [{\citenamefont {Lagos}\ \emph {et~al.}(2017)\citenamefont {Lagos},
  \citenamefont {Tr{\"u}gler}, \citenamefont {Hohenester},\ and\ \citenamefont
  {Batson}}]{lagos2017mapping}%
  \BibitemOpen
  \bibfield  {author} {\bibinfo {author} {\bibfnamefont {M.~J.}\ \bibnamefont
  {Lagos}}, \bibinfo {author} {\bibfnamefont {A.}~\bibnamefont {Tr{\"u}gler}},
  \bibinfo {author} {\bibfnamefont {U.}~\bibnamefont {Hohenester}},\ and\
  \bibinfo {author} {\bibfnamefont {P.~E.}\ \bibnamefont {Batson}},\ }\bibfield
   {title} {\bibinfo {title} {Mapping vibrational surface and bulk modes in a
  single nanocube},\ }\href@noop {} {\bibfield  {journal} {\bibinfo  {journal}
  {Nature}\ }\textbf {\bibinfo {volume} {543}},\ \bibinfo {pages} {529}
  (\bibinfo {year} {2017})}\BibitemShut {NoStop}%
\bibitem [{\citenamefont {Hage}\ \emph {et~al.}(2018)\citenamefont {Hage},
  \citenamefont {Nicholls}, \citenamefont {Yates}, \citenamefont {McCulloch},
  \citenamefont {Lovejoy}, \citenamefont {Dellby}, \citenamefont {Krivanek},
  \citenamefont {Refson},\ and\ \citenamefont {Ramasse}}]{hage2018nanoscale}%
  \BibitemOpen
  \bibfield  {author} {\bibinfo {author} {\bibfnamefont {F.~S.}\ \bibnamefont
  {Hage}}, \bibinfo {author} {\bibfnamefont {R.~J.}\ \bibnamefont {Nicholls}},
  \bibinfo {author} {\bibfnamefont {J.~R.}\ \bibnamefont {Yates}}, \bibinfo
  {author} {\bibfnamefont {D.~G.}\ \bibnamefont {McCulloch}}, \bibinfo {author}
  {\bibfnamefont {T.~C.}\ \bibnamefont {Lovejoy}}, \bibinfo {author}
  {\bibfnamefont {N.}~\bibnamefont {Dellby}}, \bibinfo {author} {\bibfnamefont
  {O.~L.}\ \bibnamefont {Krivanek}}, \bibinfo {author} {\bibfnamefont
  {K.}~\bibnamefont {Refson}},\ and\ \bibinfo {author} {\bibfnamefont {Q.~M.}\
  \bibnamefont {Ramasse}},\ }\bibfield  {title} {\bibinfo {title} {Nanoscale
  momentum-resolved vibrational spectroscopy},\ }\href@noop {} {\bibfield
  {journal} {\bibinfo  {journal} {Sci. Adv.}\ }\textbf {\bibinfo {volume}
  {4}},\ \bibinfo {pages} {eaar7495} (\bibinfo {year} {2018})}\BibitemShut
  {NoStop}%
\bibitem [{\citenamefont {Hachtel}\ \emph {et~al.}(2019)\citenamefont
  {Hachtel}, \citenamefont {Huang}, \citenamefont {Popovs}, \citenamefont
  {Jansone-Popova}, \citenamefont {Keum}, \citenamefont {Jakowski},
  \citenamefont {Lovejoy}, \citenamefont {Dellby}, \citenamefont {Krivanek},\
  and\ \citenamefont {Idrobo}}]{hachtel2019identification}%
  \BibitemOpen
  \bibfield  {author} {\bibinfo {author} {\bibfnamefont {J.~A.}\ \bibnamefont
  {Hachtel}}, \bibinfo {author} {\bibfnamefont {J.}~\bibnamefont {Huang}},
  \bibinfo {author} {\bibfnamefont {I.}~\bibnamefont {Popovs}}, \bibinfo
  {author} {\bibfnamefont {S.}~\bibnamefont {Jansone-Popova}}, \bibinfo
  {author} {\bibfnamefont {J.~K.}\ \bibnamefont {Keum}}, \bibinfo {author}
  {\bibfnamefont {J.}~\bibnamefont {Jakowski}}, \bibinfo {author}
  {\bibfnamefont {T.~C.}\ \bibnamefont {Lovejoy}}, \bibinfo {author}
  {\bibfnamefont {N.}~\bibnamefont {Dellby}}, \bibinfo {author} {\bibfnamefont
  {O.~L.}\ \bibnamefont {Krivanek}},\ and\ \bibinfo {author} {\bibfnamefont
  {J.~C.}\ \bibnamefont {Idrobo}},\ }\bibfield  {title} {\bibinfo {title}
  {Identification of site-specific isotopic labels by vibrational spectroscopy
  in the electron microscope},\ }\href@noop {} {\bibfield  {journal} {\bibinfo
  {journal} {Science}\ }\textbf {\bibinfo {volume} {363}},\ \bibinfo {pages}
  {525} (\bibinfo {year} {2019})}\BibitemShut {NoStop}%
\bibitem [{\citenamefont {Senga}\ \emph {et~al.}(2019)\citenamefont {Senga},
  \citenamefont {Suenaga}, \citenamefont {Barone}, \citenamefont {Morishita},
  \citenamefont {Mauri},\ and\ \citenamefont {Pichler}}]{senga2019position}%
  \BibitemOpen
  \bibfield  {author} {\bibinfo {author} {\bibfnamefont {R.}~\bibnamefont
  {Senga}}, \bibinfo {author} {\bibfnamefont {K.}~\bibnamefont {Suenaga}},
  \bibinfo {author} {\bibfnamefont {P.}~\bibnamefont {Barone}}, \bibinfo
  {author} {\bibfnamefont {S.}~\bibnamefont {Morishita}}, \bibinfo {author}
  {\bibfnamefont {F.}~\bibnamefont {Mauri}},\ and\ \bibinfo {author}
  {\bibfnamefont {T.}~\bibnamefont {Pichler}},\ }\bibfield  {title} {\bibinfo
  {title} {Position and momentum mapping of vibrations in graphene
  nanostructures},\ }\href@noop {} {\bibfield  {journal} {\bibinfo  {journal}
  {Nature}\ }\textbf {\bibinfo {volume} {573}},\ \bibinfo {pages} {247}
  (\bibinfo {year} {2019})}\BibitemShut {NoStop}%
\bibitem [{\citenamefont {Hong}\ \emph {et~al.}(2020)\citenamefont {Hong},
  \citenamefont {Senga}, \citenamefont {Pichler},\ and\ \citenamefont
  {Suenaga}}]{hong2020probing}%
  \BibitemOpen
  \bibfield  {author} {\bibinfo {author} {\bibfnamefont {J.}~\bibnamefont
  {Hong}}, \bibinfo {author} {\bibfnamefont {R.}~\bibnamefont {Senga}},
  \bibinfo {author} {\bibfnamefont {T.}~\bibnamefont {Pichler}},\ and\ \bibinfo
  {author} {\bibfnamefont {K.}~\bibnamefont {Suenaga}},\ }\bibfield  {title}
  {\bibinfo {title} {Probing exciton dispersions of freestanding monolayer wse
  2 by momentum-resolved electron energy-loss spectroscopy},\ }\href@noop {}
  {\bibfield  {journal} {\bibinfo  {journal} {Phys. Rev. Lett.}\ }\textbf
  {\bibinfo {volume} {124}},\ \bibinfo {pages} {087401} (\bibinfo {year}
  {2020})}\BibitemShut {NoStop}%
\bibitem [{\citenamefont {O'Hara}\ \emph {et~al.}(2023)\citenamefont {O'Hara},
  \citenamefont {Plotkin-Swing}, \citenamefont {Dellby}, \citenamefont
  {Idrobo}, \citenamefont {Krivanek}, \citenamefont {Lovejoy},\ and\
  \citenamefont {Pantelides}}]{ohara2023high}%
  \BibitemOpen
  \bibfield  {author} {\bibinfo {author} {\bibfnamefont {A.}~\bibnamefont
  {O'Hara}}, \bibinfo {author} {\bibfnamefont {B.}~\bibnamefont
  {Plotkin-Swing}}, \bibinfo {author} {\bibfnamefont {N.}~\bibnamefont
  {Dellby}}, \bibinfo {author} {\bibfnamefont {J.~C.}\ \bibnamefont {Idrobo}},
  \bibinfo {author} {\bibfnamefont {O.~L.}\ \bibnamefont {Krivanek}}, \bibinfo
  {author} {\bibfnamefont {T.~C.}\ \bibnamefont {Lovejoy}},\ and\ \bibinfo
  {author} {\bibfnamefont {S.~T.}\ \bibnamefont {Pantelides}},\ }\bibfield
  {title} {\bibinfo {title} {High-temperature phonons in h-{BN}:
  momentum-resolved vibrational spectroscopy and theory},\ }\href@noop {}
  {\bibfield  {journal} {\bibinfo  {journal} {arXiv preprint arXiv:2310.13813}\
  } (\bibinfo {year} {2023})}\BibitemShut {NoStop}%
\bibitem [{\citenamefont {Li}\ \emph {et~al.}(2023)\citenamefont {Li},
  \citenamefont {Li}, \citenamefont {Tang}, \citenamefont {Tao}, \citenamefont
  {Xue}, \citenamefont {Liu}, \citenamefont {Peng}, \citenamefont {Chen},
  \citenamefont {Guo},\ and\ \citenamefont {Zhu}}]{li2023direct}%
  \BibitemOpen
  \bibfield  {author} {\bibinfo {author} {\bibfnamefont {J.}~\bibnamefont
  {Li}}, \bibinfo {author} {\bibfnamefont {J.}~\bibnamefont {Li}}, \bibinfo
  {author} {\bibfnamefont {J.}~\bibnamefont {Tang}}, \bibinfo {author}
  {\bibfnamefont {Z.}~\bibnamefont {Tao}}, \bibinfo {author} {\bibfnamefont
  {S.}~\bibnamefont {Xue}}, \bibinfo {author} {\bibfnamefont {J.}~\bibnamefont
  {Liu}}, \bibinfo {author} {\bibfnamefont {H.}~\bibnamefont {Peng}}, \bibinfo
  {author} {\bibfnamefont {X.-Q.}\ \bibnamefont {Chen}}, \bibinfo {author}
  {\bibfnamefont {J.}~\bibnamefont {Guo}},\ and\ \bibinfo {author}
  {\bibfnamefont {X.}~\bibnamefont {Zhu}},\ }\bibfield  {title} {\bibinfo
  {title} {Direct observation of topological phonons in graphene},\ }\href@noop
  {} {\bibfield  {journal} {\bibinfo  {journal} {Phys. Rev. Lett.}\ }\textbf
  {\bibinfo {volume} {131}},\ \bibinfo {pages} {116602} (\bibinfo {year}
  {2023})}\BibitemShut {NoStop}%
\bibitem [{\citenamefont {Li}\ \emph {et~al.}(2024)\citenamefont {Li},
  \citenamefont {Wang}, \citenamefont {Wang}, \citenamefont {Tao},
  \citenamefont {Zhong}, \citenamefont {Su}, \citenamefont {Xue}, \citenamefont
  {Miao}, \citenamefont {Wang}, \citenamefont {Peng}, \citenamefont {Guo},\
  and\ \citenamefont {Zhu}}]{li2024observation}%
  \BibitemOpen
  \bibfield  {author} {\bibinfo {author} {\bibfnamefont {J.}~\bibnamefont
  {Li}}, \bibinfo {author} {\bibfnamefont {L.}~\bibnamefont {Wang}}, \bibinfo
  {author} {\bibfnamefont {Y.}~\bibnamefont {Wang}}, \bibinfo {author}
  {\bibfnamefont {Z.}~\bibnamefont {Tao}}, \bibinfo {author} {\bibfnamefont
  {W.}~\bibnamefont {Zhong}}, \bibinfo {author} {\bibfnamefont
  {Z.}~\bibnamefont {Su}}, \bibinfo {author} {\bibfnamefont {S.}~\bibnamefont
  {Xue}}, \bibinfo {author} {\bibfnamefont {G.}~\bibnamefont {Miao}}, \bibinfo
  {author} {\bibfnamefont {W.}~\bibnamefont {Wang}}, \bibinfo {author}
  {\bibfnamefont {H.}~\bibnamefont {Peng}}, \bibinfo {author} {\bibfnamefont
  {J.}~\bibnamefont {Guo}},\ and\ \bibinfo {author} {\bibfnamefont
  {X.}~\bibnamefont {Zhu}},\ }\bibfield  {title} {\bibinfo {title} {Observation
  of the nonanalytic behavior of optical phonons in monolayer hexagonal boron
  nitride},\ }\href@noop {} {\bibfield  {journal} {\bibinfo  {journal} {Nat.
  Commun.}\ }\textbf {\bibinfo {volume} {15}},\ \bibinfo {pages} {1938}
  (\bibinfo {year} {2024})}\BibitemShut {NoStop}%
\bibitem [{\citenamefont {Yuan}\ and\ \citenamefont
  {Menon}(1997)}]{Yuan1997-hj}%
  \BibitemOpen
  \bibfield  {author} {\bibinfo {author} {\bibfnamefont {J.}~\bibnamefont
  {Yuan}}\ and\ \bibinfo {author} {\bibfnamefont {N.~K.}\ \bibnamefont
  {Menon}},\ }\bibfield  {title} {\bibinfo {title} {Magnetic linear dichroism
  in electron energy loss spectroscopy},\ }\href@noop {} {\bibfield  {journal}
  {\bibinfo  {journal} {J. Appl. Phys.}\ }\textbf {\bibinfo {volume} {81}},\
  \bibinfo {pages} {5087} (\bibinfo {year} {1997})}\BibitemShut {NoStop}%
\bibitem [{\citenamefont {H{\'e}bert}\ and\ \citenamefont
  {Schattschneider}(2003)}]{Hebert2003-mn}%
  \BibitemOpen
  \bibfield  {author} {\bibinfo {author} {\bibfnamefont {C.}~\bibnamefont
  {H{\'e}bert}}\ and\ \bibinfo {author} {\bibfnamefont {P.}~\bibnamefont
  {Schattschneider}},\ }\bibfield  {title} {\bibinfo {title} {A proposal for
  dichroic experiments in the electron microscope},\ }\href@noop {} {\bibfield
  {journal} {\bibinfo  {journal} {Ultramicroscopy}\ }\textbf {\bibinfo {volume}
  {96}},\ \bibinfo {pages} {463} (\bibinfo {year} {2003})}\BibitemShut
  {NoStop}%
\bibitem [{\citenamefont {Schattschneider}\ \emph {et~al.}(2006)\citenamefont
  {Schattschneider}, \citenamefont {Rubino}, \citenamefont {H{\'e}bert},
  \citenamefont {Rusz}, \citenamefont {Kune{\v s}}, \citenamefont {Nov{\'a}k},
  \citenamefont {Carlino}, \citenamefont {Fabrizioli}, \citenamefont
  {Panaccione},\ and\ \citenamefont {Rossi}}]{Schattschneider2006-fy}%
  \BibitemOpen
  \bibfield  {author} {\bibinfo {author} {\bibfnamefont {P.}~\bibnamefont
  {Schattschneider}}, \bibinfo {author} {\bibfnamefont {S.}~\bibnamefont
  {Rubino}}, \bibinfo {author} {\bibfnamefont {C.}~\bibnamefont {H{\'e}bert}},
  \bibinfo {author} {\bibfnamefont {J.}~\bibnamefont {Rusz}}, \bibinfo {author}
  {\bibfnamefont {J.}~\bibnamefont {Kune{\v s}}}, \bibinfo {author}
  {\bibfnamefont {P.}~\bibnamefont {Nov{\'a}k}}, \bibinfo {author}
  {\bibfnamefont {E.}~\bibnamefont {Carlino}}, \bibinfo {author} {\bibfnamefont
  {M.}~\bibnamefont {Fabrizioli}}, \bibinfo {author} {\bibfnamefont
  {G.}~\bibnamefont {Panaccione}},\ and\ \bibinfo {author} {\bibfnamefont
  {G.}~\bibnamefont {Rossi}},\ }\bibfield  {title} {\bibinfo {title} {Detection
  of magnetic circular dichroism using a transmission electron microscope},\
  }\href@noop {} {\bibfield  {journal} {\bibinfo  {journal} {Nature}\ }\textbf
  {\bibinfo {volume} {441}},\ \bibinfo {pages} {486} (\bibinfo {year}
  {2006})}\BibitemShut {NoStop}%
\bibitem [{\citenamefont {Guzzinati}\ \emph {et~al.}(2017)\citenamefont
  {Guzzinati}, \citenamefont {B{\'e}ch{\'e}}, \citenamefont
  {Louren{\c{c}}o-Martins}, \citenamefont {Martin}, \citenamefont {Kociak},\
  and\ \citenamefont {Verbeeck}}]{guzzinati2017probing}%
  \BibitemOpen
  \bibfield  {author} {\bibinfo {author} {\bibfnamefont {G.}~\bibnamefont
  {Guzzinati}}, \bibinfo {author} {\bibfnamefont {A.}~\bibnamefont
  {B{\'e}ch{\'e}}}, \bibinfo {author} {\bibfnamefont {H.}~\bibnamefont
  {Louren{\c{c}}o-Martins}}, \bibinfo {author} {\bibfnamefont {J.}~\bibnamefont
  {Martin}}, \bibinfo {author} {\bibfnamefont {M.}~\bibnamefont {Kociak}},\
  and\ \bibinfo {author} {\bibfnamefont {J.}~\bibnamefont {Verbeeck}},\
  }\bibfield  {title} {\bibinfo {title} {Probing the symmetry of the potential
  of localized surface plasmon resonances with phase-shaped electron beams},\
  }\href@noop {} {\bibfield  {journal} {\bibinfo  {journal} {Nat. Commun.}\
  }\textbf {\bibinfo {volume} {8}},\ \bibinfo {pages} {1} (\bibinfo {year}
  {2017})}\BibitemShut {NoStop}%
\bibitem [{\citenamefont {Asenjo-Garcia}\ and\ \citenamefont {Garc\'ia~de
  Abajo}(2014)}]{asenjo2014dichroism}%
  \BibitemOpen
  \bibfield  {author} {\bibinfo {author} {\bibfnamefont {A.}~\bibnamefont
  {Asenjo-Garcia}}\ and\ \bibinfo {author} {\bibfnamefont {F.~J.}\ \bibnamefont
  {Garc\'ia~de Abajo}},\ }\bibfield  {title} {\bibinfo {title} {Dichroism in
  the interaction between vortex electron beams, plasmons, and molecules},\
  }\href@noop {} {\bibfield  {journal} {\bibinfo  {journal} {Phys. Rev. Lett.}\
  }\textbf {\bibinfo {volume} {113}},\ \bibinfo {pages} {066102} (\bibinfo
  {year} {2014})}\BibitemShut {NoStop}%
\bibitem [{\citenamefont {Zanfrognini}\ \emph {et~al.}(2019)\citenamefont
  {Zanfrognini}, \citenamefont {Rotunno}, \citenamefont {Frabboni},
  \citenamefont {Sit}, \citenamefont {Karimi}, \citenamefont {Hohenester},\
  and\ \citenamefont {Grillo}}]{zanfrognini2019orbital}%
  \BibitemOpen
  \bibfield  {author} {\bibinfo {author} {\bibfnamefont {M.}~\bibnamefont
  {Zanfrognini}}, \bibinfo {author} {\bibfnamefont {E.}~\bibnamefont
  {Rotunno}}, \bibinfo {author} {\bibfnamefont {S.}~\bibnamefont {Frabboni}},
  \bibinfo {author} {\bibfnamefont {A.}~\bibnamefont {Sit}}, \bibinfo {author}
  {\bibfnamefont {E.}~\bibnamefont {Karimi}}, \bibinfo {author} {\bibfnamefont
  {U.}~\bibnamefont {Hohenester}},\ and\ \bibinfo {author} {\bibfnamefont
  {V.}~\bibnamefont {Grillo}},\ }\bibfield  {title} {\bibinfo {title} {Orbital
  angular momentum and energy loss characterization of plasmonic excitations in
  metallic nanostructures in {TEM}},\ }\href@noop {} {\bibfield  {journal}
  {\bibinfo  {journal} {ACS Photonics}\ }\textbf {\bibinfo {volume} {6}},\
  \bibinfo {pages} {620} (\bibinfo {year} {2019})}\BibitemShut {NoStop}%
\bibitem [{\citenamefont {Louren{\c{c}}o-Martins}\ \emph
  {et~al.}(2021)\citenamefont {Louren{\c{c}}o-Martins}, \citenamefont
  {G{\'e}rard},\ and\ \citenamefont {Kociak}}]{lourencco2021optical}%
  \BibitemOpen
  \bibfield  {author} {\bibinfo {author} {\bibfnamefont {H.}~\bibnamefont
  {Louren{\c{c}}o-Martins}}, \bibinfo {author} {\bibfnamefont {D.}~\bibnamefont
  {G{\'e}rard}},\ and\ \bibinfo {author} {\bibfnamefont {M.}~\bibnamefont
  {Kociak}},\ }\bibfield  {title} {\bibinfo {title} {Optical polarization
  analogue in free electron beams},\ }\href@noop {} {\bibfield  {journal}
  {\bibinfo  {journal} {Nat. Phys.}\ }\textbf {\bibinfo {volume} {17}},\
  \bibinfo {pages} {598} (\bibinfo {year} {2021})}\BibitemShut {NoStop}%
\bibitem [{\citenamefont {Bourgeois}\ \emph {et~al.}(2023)\citenamefont
  {Bourgeois}, \citenamefont {Nixon}, \citenamefont {Chalifour},\ and\
  \citenamefont {Masiello}}]{bourgeois2023optical}%
  \BibitemOpen
  \bibfield  {author} {\bibinfo {author} {\bibfnamefont {M.~R.}\ \bibnamefont
  {Bourgeois}}, \bibinfo {author} {\bibfnamefont {A.~G.}\ \bibnamefont
  {Nixon}}, \bibinfo {author} {\bibfnamefont {M.}~\bibnamefont {Chalifour}},\
  and\ \bibinfo {author} {\bibfnamefont {D.~J.}\ \bibnamefont {Masiello}},\
  }\bibfield  {title} {\bibinfo {title} {Optical polarization analogs in
  inelastic free-electron scattering},\ }\href
  {https://doi.org/10.1126/sciadv.adj6038} {\bibfield  {journal} {\bibinfo
  {journal} {Sci. Adv.}\ }\textbf {\bibinfo {volume} {9}},\ \bibinfo {pages}
  {eadj6038} (\bibinfo {year} {2023})}\BibitemShut {NoStop}%
\bibitem [{\citenamefont {Niermann}\ \emph {et~al.}(2014)\citenamefont
  {Niermann}, \citenamefont {Verbeeck},\ and\ \citenamefont
  {Lehmann}}]{niermann2014creating}%
  \BibitemOpen
  \bibfield  {author} {\bibinfo {author} {\bibfnamefont {T.}~\bibnamefont
  {Niermann}}, \bibinfo {author} {\bibfnamefont {J.}~\bibnamefont {Verbeeck}},\
  and\ \bibinfo {author} {\bibfnamefont {M.}~\bibnamefont {Lehmann}},\
  }\bibfield  {title} {\bibinfo {title} {Creating arrays of electron
  vortices},\ }\href@noop {} {\bibfield  {journal} {\bibinfo  {journal}
  {Ultramicroscopy}\ }\textbf {\bibinfo {volume} {136}},\ \bibinfo {pages}
  {165} (\bibinfo {year} {2014})}\BibitemShut {NoStop}%
\bibitem [{\citenamefont {Yu}\ \emph {et~al.}(2023)\citenamefont {Yu},
  \citenamefont {Vega~Iba{\~n}ez}, \citenamefont {B{\'e}ch{\'e}},\ and\
  \citenamefont {Verbeeck}}]{yu2023quantum}%
  \BibitemOpen
  \bibfield  {author} {\bibinfo {author} {\bibfnamefont {C.-P.}\ \bibnamefont
  {Yu}}, \bibinfo {author} {\bibfnamefont {F.}~\bibnamefont {Vega~Iba{\~n}ez}},
  \bibinfo {author} {\bibfnamefont {A.}~\bibnamefont {B{\'e}ch{\'e}}},\ and\
  \bibinfo {author} {\bibfnamefont {J.}~\bibnamefont {Verbeeck}},\ }\bibfield
  {title} {\bibinfo {title} {Quantum wavefront shaping with a 48-element
  programmable phase plate for electrons},\ }\href@noop {} {\bibfield
  {journal} {\bibinfo  {journal} {SciPost Phys.}\ }\textbf {\bibinfo {volume}
  {15}},\ \bibinfo {pages} {223} (\bibinfo {year} {2023})}\BibitemShut
  {NoStop}%
\bibitem [{\citenamefont {Feist}\ \emph {et~al.}(2020)\citenamefont {Feist},
  \citenamefont {Yalunin}, \citenamefont {Sch\"afer},\ and\ \citenamefont
  {Ropers}}]{PhysRevResearch.2.043227}%
  \BibitemOpen
  \bibfield  {author} {\bibinfo {author} {\bibfnamefont {A.}~\bibnamefont
  {Feist}}, \bibinfo {author} {\bibfnamefont {S.~V.}\ \bibnamefont {Yalunin}},
  \bibinfo {author} {\bibfnamefont {S.}~\bibnamefont {Sch\"afer}},\ and\
  \bibinfo {author} {\bibfnamefont {C.}~\bibnamefont {Ropers}},\ }\bibfield
  {title} {\bibinfo {title} {High-purity free-electron momentum states prepared
  by three-dimensional optical phase modulation},\ }\href
  {https://doi.org/10.1103/PhysRevResearch.2.043227} {\bibfield  {journal}
  {\bibinfo  {journal} {Phys. Rev. Res.}\ }\textbf {\bibinfo {volume} {2}},\
  \bibinfo {pages} {043227} (\bibinfo {year} {2020})}\BibitemShut {NoStop}%
\bibitem [{\citenamefont {Madan}\ \emph {et~al.}(2022)\citenamefont {Madan},
  \citenamefont {Leccese}, \citenamefont {Mazur}, \citenamefont {Barantani},
  \citenamefont {LaGrange}, \citenamefont {Sapozhnik}, \citenamefont {Tengdin},
  \citenamefont {Gargiulo}, \citenamefont {Rotunno}, \citenamefont {Olaya},
  \citenamefont {Kaminer}, \citenamefont {Grillo}, \citenamefont
  {Garc{\'\i}a~de Abajo}, \citenamefont {Carbone},\ and\ \citenamefont
  {Vanacore}}]{Madan2022-gj}%
  \BibitemOpen
  \bibfield  {author} {\bibinfo {author} {\bibfnamefont {I.}~\bibnamefont
  {Madan}}, \bibinfo {author} {\bibfnamefont {V.}~\bibnamefont {Leccese}},
  \bibinfo {author} {\bibfnamefont {A.}~\bibnamefont {Mazur}}, \bibinfo
  {author} {\bibfnamefont {F.}~\bibnamefont {Barantani}}, \bibinfo {author}
  {\bibfnamefont {T.}~\bibnamefont {LaGrange}}, \bibinfo {author}
  {\bibfnamefont {A.}~\bibnamefont {Sapozhnik}}, \bibinfo {author}
  {\bibfnamefont {P.~M.}\ \bibnamefont {Tengdin}}, \bibinfo {author}
  {\bibfnamefont {S.}~\bibnamefont {Gargiulo}}, \bibinfo {author}
  {\bibfnamefont {E.}~\bibnamefont {Rotunno}}, \bibinfo {author} {\bibfnamefont
  {J.-C.}\ \bibnamefont {Olaya}}, \bibinfo {author} {\bibfnamefont
  {I.}~\bibnamefont {Kaminer}}, \bibinfo {author} {\bibfnamefont
  {V.}~\bibnamefont {Grillo}}, \bibinfo {author} {\bibfnamefont {F.~J.}\
  \bibnamefont {Garc{\'\i}a~de Abajo}}, \bibinfo {author} {\bibfnamefont
  {F.}~\bibnamefont {Carbone}},\ and\ \bibinfo {author} {\bibfnamefont {G.~M.}\
  \bibnamefont {Vanacore}},\ }\bibfield  {title} {\bibinfo {title} {Ultrafast
  transverse modulation of free electrons by interaction with shaped optical
  fields},\ }\href@noop {} {\bibfield  {journal} {\bibinfo  {journal} {ACS
  Photonics}\ }\textbf {\bibinfo {volume} {9}},\ \bibinfo {pages} {3215}
  (\bibinfo {year} {2022})}\BibitemShut {NoStop}%
\bibitem [{\citenamefont {Harvey}\ \emph {et~al.}(2014)\citenamefont {Harvey},
  \citenamefont {Pierce}, \citenamefont {Agrawal}, \citenamefont {Ercius},
  \citenamefont {Linck},\ and\ \citenamefont {McMorran}}]{harvey2014efficient}%
  \BibitemOpen
  \bibfield  {author} {\bibinfo {author} {\bibfnamefont {T.~R.}\ \bibnamefont
  {Harvey}}, \bibinfo {author} {\bibfnamefont {J.~S.}\ \bibnamefont {Pierce}},
  \bibinfo {author} {\bibfnamefont {A.~K.}\ \bibnamefont {Agrawal}}, \bibinfo
  {author} {\bibfnamefont {P.}~\bibnamefont {Ercius}}, \bibinfo {author}
  {\bibfnamefont {M.}~\bibnamefont {Linck}},\ and\ \bibinfo {author}
  {\bibfnamefont {B.~J.}\ \bibnamefont {McMorran}},\ }\bibfield  {title}
  {\bibinfo {title} {Efficient diffractive phase optics for electrons},\
  }\href@noop {} {\bibfield  {journal} {\bibinfo  {journal} {New J. Phys.}\
  }\textbf {\bibinfo {volume} {16}},\ \bibinfo {pages} {093039} (\bibinfo
  {year} {2014})}\BibitemShut {NoStop}%
\bibitem [{\citenamefont {Tsesses}\ \emph {et~al.}(2023)\citenamefont
  {Tsesses}, \citenamefont {Dahan}, \citenamefont {Wang}, \citenamefont
  {Bucher}, \citenamefont {Cohen}, \citenamefont {Reinhardt}, \citenamefont
  {Bartal},\ and\ \citenamefont {Kaminer}}]{Tsesses2023-hw}%
  \BibitemOpen
  \bibfield  {author} {\bibinfo {author} {\bibfnamefont {S.}~\bibnamefont
  {Tsesses}}, \bibinfo {author} {\bibfnamefont {R.}~\bibnamefont {Dahan}},
  \bibinfo {author} {\bibfnamefont {K.}~\bibnamefont {Wang}}, \bibinfo {author}
  {\bibfnamefont {T.}~\bibnamefont {Bucher}}, \bibinfo {author} {\bibfnamefont
  {K.}~\bibnamefont {Cohen}}, \bibinfo {author} {\bibfnamefont
  {O.}~\bibnamefont {Reinhardt}}, \bibinfo {author} {\bibfnamefont
  {G.}~\bibnamefont {Bartal}},\ and\ \bibinfo {author} {\bibfnamefont
  {I.}~\bibnamefont {Kaminer}},\ }\bibfield  {title} {\bibinfo {title} {Tunable
  photon-induced spatial modulation of free electrons},\ }\href@noop {}
  {\bibfield  {journal} {\bibinfo  {journal} {Nat. Mater.}\ }\textbf {\bibinfo
  {volume} {22}},\ \bibinfo {pages} {345} (\bibinfo {year} {2023})}\BibitemShut
  {NoStop}%
\bibitem [{\citenamefont {Lloyd}\ \emph {et~al.}(2012)\citenamefont {Lloyd},
  \citenamefont {Babiker},\ and\ \citenamefont {Yuan}}]{lloyd2012quantized}%
  \BibitemOpen
  \bibfield  {author} {\bibinfo {author} {\bibfnamefont {S.}~\bibnamefont
  {Lloyd}}, \bibinfo {author} {\bibfnamefont {M.}~\bibnamefont {Babiker}},\
  and\ \bibinfo {author} {\bibfnamefont {J.}~\bibnamefont {Yuan}},\ }\bibfield
  {title} {\bibinfo {title} {Quantized orbital angular momentum transfer and
  magnetic dichroism in the interaction of electron vortices with matter},\
  }\href@noop {} {\bibfield  {journal} {\bibinfo  {journal} {Phys. Rev. Lett.}\
  }\textbf {\bibinfo {volume} {108}},\ \bibinfo {pages} {074802} (\bibinfo
  {year} {2012})}\BibitemShut {NoStop}%
\bibitem [{\citenamefont {Cai}\ \emph {et~al.}(2018)\citenamefont {Cai},
  \citenamefont {Reinhardt}, \citenamefont {Kaminer},\ and\ \citenamefont
  {Garc\'ia~de Abajo}}]{cai2018efficient}%
  \BibitemOpen
  \bibfield  {author} {\bibinfo {author} {\bibfnamefont {W.}~\bibnamefont
  {Cai}}, \bibinfo {author} {\bibfnamefont {O.}~\bibnamefont {Reinhardt}},
  \bibinfo {author} {\bibfnamefont {I.}~\bibnamefont {Kaminer}},\ and\ \bibinfo
  {author} {\bibfnamefont {F.~J.}\ \bibnamefont {Garc\'ia~de Abajo}},\
  }\bibfield  {title} {\bibinfo {title} {Efficient orbital angular momentum
  transfer between plasmons and free electrons},\ }\href@noop {} {\bibfield
  {journal} {\bibinfo  {journal} {Phys. Rev. B}\ }\textbf {\bibinfo {volume}
  {98}},\ \bibinfo {pages} {045424} (\bibinfo {year} {2018})}\BibitemShut
  {NoStop}%
\bibitem [{\citenamefont {Bourgeois}\ \emph {et~al.}(2022)\citenamefont
  {Bourgeois}, \citenamefont {Nixon}, \citenamefont {Chalifour}, \citenamefont
  {Beutler},\ and\ \citenamefont {Masiello}}]{bourgeois2022polarization}%
  \BibitemOpen
  \bibfield  {author} {\bibinfo {author} {\bibfnamefont {M.~R.}\ \bibnamefont
  {Bourgeois}}, \bibinfo {author} {\bibfnamefont {A.~G.}\ \bibnamefont
  {Nixon}}, \bibinfo {author} {\bibfnamefont {M.}~\bibnamefont {Chalifour}},
  \bibinfo {author} {\bibfnamefont {E.~K.}\ \bibnamefont {Beutler}},\ and\
  \bibinfo {author} {\bibfnamefont {D.~J.}\ \bibnamefont {Masiello}},\
  }\bibfield  {title} {\bibinfo {title} {Polarization-resolved electron energy
  gain nanospectroscopy with phase-structured electron beams},\ }\href@noop {}
  {\bibfield  {journal} {\bibinfo  {journal} {Nano Lett.}\ }\textbf {\bibinfo
  {volume} {22}},\ \bibinfo {pages} {7158} (\bibinfo {year}
  {2022})}\BibitemShut {NoStop}%
\bibitem [{\citenamefont {Dove}(1993)}]{dove1993lattice}%
  \BibitemOpen
  \bibfield  {author} {\bibinfo {author} {\bibfnamefont {M.~T.}\ \bibnamefont
  {Dove}},\ }\href@noop {} {\emph {\bibinfo {title} {Introduction to Lattice
  Dynamics}}}\ (\bibinfo  {publisher} {Cambridge University Press},\ \bibinfo
  {year} {1993})\BibitemShut {NoStop}%
\bibitem [{\citenamefont {Bosak}\ \emph {et~al.}(2007)\citenamefont {Bosak},
  \citenamefont {Krisch}, \citenamefont {Mohr}, \citenamefont {Maultzsch},\
  and\ \citenamefont {Thomsen}}]{bosak2007elasticity}%
  \BibitemOpen
  \bibfield  {author} {\bibinfo {author} {\bibfnamefont {A.}~\bibnamefont
  {Bosak}}, \bibinfo {author} {\bibfnamefont {M.}~\bibnamefont {Krisch}},
  \bibinfo {author} {\bibfnamefont {M.}~\bibnamefont {Mohr}}, \bibinfo {author}
  {\bibfnamefont {J.}~\bibnamefont {Maultzsch}},\ and\ \bibinfo {author}
  {\bibfnamefont {C.}~\bibnamefont {Thomsen}},\ }\bibfield  {title} {\bibinfo
  {title} {Elasticity of single-crystalline graphite: Inelastic x-ray
  scattering study},\ }\href@noop {} {\bibfield  {journal} {\bibinfo  {journal}
  {Phys. Rev. B}\ }\textbf {\bibinfo {volume} {75}},\ \bibinfo {pages} {153408}
  (\bibinfo {year} {2007})}\BibitemShut {NoStop}%
\bibitem [{\citenamefont {Michel}\ and\ \citenamefont
  {Verberck}(2008)}]{michel2008theory}%
  \BibitemOpen
  \bibfield  {author} {\bibinfo {author} {\bibfnamefont {K.}~\bibnamefont
  {Michel}}\ and\ \bibinfo {author} {\bibfnamefont {B.}~\bibnamefont
  {Verberck}},\ }\bibfield  {title} {\bibinfo {title} {Theory of the evolution
  of phonon spectra and elastic constants from graphene to graphite},\
  }\href@noop {} {\bibfield  {journal} {\bibinfo  {journal} {Phys. Rev. B}\
  }\textbf {\bibinfo {volume} {78}},\ \bibinfo {pages} {085424} (\bibinfo
  {year} {2008})}\BibitemShut {NoStop}%
\bibitem [{\citenamefont {Michel}\ and\ \citenamefont
  {Verberck}(2009)}]{michel2009theory}%
  \BibitemOpen
  \bibfield  {author} {\bibinfo {author} {\bibfnamefont {K.}~\bibnamefont
  {Michel}}\ and\ \bibinfo {author} {\bibfnamefont {B.}~\bibnamefont
  {Verberck}},\ }\bibfield  {title} {\bibinfo {title} {Theory of elastic and
  piezoelectric effects in two-dimensional hexagonal boron nitride},\
  }\href@noop {} {\bibfield  {journal} {\bibinfo  {journal} {Phys. Rev. B}\
  }\textbf {\bibinfo {volume} {80}},\ \bibinfo {pages} {224301} (\bibinfo
  {year} {2009})}\BibitemShut {NoStop}%
\bibitem [{SM_()}]{SM_chiralphonon}%
  \BibitemOpen
  \href@noop {} {}\bibinfo {note} {Supplemental Material at [url] includes a
  description of the employed harmonic crystal hBN and graphene phonon models,
  derivations of main equations, spatial maps of PAM-resolved EELS, and
  pseudoangular momentum considerations.}\BibitemShut {Stop}%
\bibitem [{\citenamefont {Hohenester}\ \emph {et~al.}(2018)\citenamefont
  {Hohenester}, \citenamefont {Tr{\"u}gler}, \citenamefont {Batson},\ and\
  \citenamefont {Lagos}}]{hohenester2018inelastic}%
  \BibitemOpen
  \bibfield  {author} {\bibinfo {author} {\bibfnamefont {U.}~\bibnamefont
  {Hohenester}}, \bibinfo {author} {\bibfnamefont {A.}~\bibnamefont
  {Tr{\"u}gler}}, \bibinfo {author} {\bibfnamefont {P.~E.}\ \bibnamefont
  {Batson}},\ and\ \bibinfo {author} {\bibfnamefont {M.~J.}\ \bibnamefont
  {Lagos}},\ }\bibfield  {title} {\bibinfo {title} {Inelastic vibrational bulk
  and surface losses of swift electrons in ionic nanostructures},\ }\href@noop
  {} {\bibfield  {journal} {\bibinfo  {journal} {Phys. Rev. B}\ }\textbf
  {\bibinfo {volume} {97}},\ \bibinfo {pages} {165418} (\bibinfo {year}
  {2018})}\BibitemShut {NoStop}%
\bibitem [{\citenamefont {Sakurai}(1985)}]{SakuraiModern}%
  \BibitemOpen
  \bibfield  {author} {\bibinfo {author} {\bibfnamefont {J.~J.}\ \bibnamefont
  {Sakurai}},\ }\href@noop {} {\emph {\bibinfo {title} {{Modern Quantum
  Mechanics}}}}\ (\bibinfo  {publisher} {Benjamin/Cummings Publishing
  Company},\ \bibinfo {year} {1985})\BibitemShut {NoStop}%
\bibitem [{\citenamefont {Rossi}\ \emph {et~al.}(2024)\citenamefont {Rossi},
  \citenamefont {Bourgeois}, \citenamefont {Walton},\ and\ \citenamefont
  {Masiello}}]{rossi2024probing}%
  \BibitemOpen
  \bibfield  {author} {\bibinfo {author} {\bibfnamefont {A.~W.}\ \bibnamefont
  {Rossi}}, \bibinfo {author} {\bibfnamefont {M.~R.}\ \bibnamefont
  {Bourgeois}}, \bibinfo {author} {\bibfnamefont {C.}~\bibnamefont {Walton}},\
  and\ \bibinfo {author} {\bibfnamefont {D.~J.}\ \bibnamefont {Masiello}},\
  }\bibfield  {title} {\bibinfo {title} {Probing the polarization of low-energy
  excitations in 2{D} materials from atomic crystals to nanophotonic arrays
  using momentum-resolved electron energy loss spectroscopy},\ }\href@noop {}
  {\bibfield  {journal} {\bibinfo  {journal} {Nano Lett.}\ }\textbf {\bibinfo
  {volume} {24}},\ \bibinfo {pages} {7748} (\bibinfo {year}
  {2024})}\BibitemShut {NoStop}%
\bibitem [{\citenamefont {Uchida}\ and\ \citenamefont
  {Tonomura}(2010)}]{uchida2010generation}%
  \BibitemOpen
  \bibfield  {author} {\bibinfo {author} {\bibfnamefont {M.}~\bibnamefont
  {Uchida}}\ and\ \bibinfo {author} {\bibfnamefont {A.}~\bibnamefont
  {Tonomura}},\ }\bibfield  {title} {\bibinfo {title} {Generation of electron
  beams carrying orbital angular momentum},\ }\href@noop {} {\bibfield
  {journal} {\bibinfo  {journal} {Nature}\ }\textbf {\bibinfo {volume} {464}},\
  \bibinfo {pages} {737} (\bibinfo {year} {2010})}\BibitemShut {NoStop}%
\bibitem [{\citenamefont {Verbeeck}\ \emph {et~al.}(2010)\citenamefont
  {Verbeeck}, \citenamefont {Tian},\ and\ \citenamefont
  {Schattschneider}}]{verbeeck2010production}%
  \BibitemOpen
  \bibfield  {author} {\bibinfo {author} {\bibfnamefont {J.}~\bibnamefont
  {Verbeeck}}, \bibinfo {author} {\bibfnamefont {H.}~\bibnamefont {Tian}},\
  and\ \bibinfo {author} {\bibfnamefont {P.}~\bibnamefont {Schattschneider}},\
  }\bibfield  {title} {\bibinfo {title} {Production and application of electron
  vortex beams},\ }\href@noop {} {\bibfield  {journal} {\bibinfo  {journal}
  {Nature}\ }\textbf {\bibinfo {volume} {467}},\ \bibinfo {pages} {301}
  (\bibinfo {year} {2010})}\BibitemShut {NoStop}%
\bibitem [{\citenamefont {Bliokh}\ \emph {et~al.}(2017)\citenamefont {Bliokh},
  \citenamefont {Ivanov}, \citenamefont {Guzzinati}, \citenamefont {Clark},
  \citenamefont {Van~Boxem}, \citenamefont {B{\'e}ch{\'e}}, \citenamefont
  {Juchtmans}, \citenamefont {Alonso}, \citenamefont {Schattschneider},
  \citenamefont {Nori},\ and\ \citenamefont {Verbeeck}}]{bliokh2017theory}%
  \BibitemOpen
  \bibfield  {author} {\bibinfo {author} {\bibfnamefont {K.~Y.}\ \bibnamefont
  {Bliokh}}, \bibinfo {author} {\bibfnamefont {I.~P.}\ \bibnamefont {Ivanov}},
  \bibinfo {author} {\bibfnamefont {G.}~\bibnamefont {Guzzinati}}, \bibinfo
  {author} {\bibfnamefont {L.}~\bibnamefont {Clark}}, \bibinfo {author}
  {\bibfnamefont {R.}~\bibnamefont {Van~Boxem}}, \bibinfo {author}
  {\bibfnamefont {A.}~\bibnamefont {B{\'e}ch{\'e}}}, \bibinfo {author}
  {\bibfnamefont {R.}~\bibnamefont {Juchtmans}}, \bibinfo {author}
  {\bibfnamefont {M.~A.}\ \bibnamefont {Alonso}}, \bibinfo {author}
  {\bibfnamefont {P.}~\bibnamefont {Schattschneider}}, \bibinfo {author}
  {\bibfnamefont {F.}~\bibnamefont {Nori}},\ and\ \bibinfo {author}
  {\bibfnamefont {J.}~\bibnamefont {Verbeeck}},\ }\bibfield  {title} {\bibinfo
  {title} {Theory and applications of free-electron vortex states},\
  }\href@noop {} {\bibfield  {journal} {\bibinfo  {journal} {Phys. Rep.}\
  }\textbf {\bibinfo {volume} {690}},\ \bibinfo {pages} {1} (\bibinfo {year}
  {2017})}\BibitemShut {NoStop}%
\end{thebibliography}%
\end{document}